\PassOptionsToPackage{unicode}{hyperref}
\PassOptionsToPackage{hyphens}{url}
\PassOptionsToPackage{dvipsnames,svgnames,x11names}{xcolor}
\documentclass[
  11pt,
]{article}
\usepackage{xcolor}
\usepackage[margin=1in]{geometry}
\usepackage{amsmath,amssymb}
\usepackage{iftex}
\ifPDFTeX
  \usepackage[T1]{fontenc}
  \usepackage[utf8]{inputenc}
  \usepackage{textcomp} 
\else 
  \usepackage{unicode-math} 
  \defaultfontfeatures{Scale=MatchLowercase}
  \defaultfontfeatures[\rmfamily]{Ligatures=TeX,Scale=1}
\fi
\usepackage{lmodern}
\ifPDFTeX\else
\fi
\IfFileExists{upquote.sty}{\usepackage{upquote}}{}
\IfFileExists{microtype.sty}{
  \usepackage[]{microtype}
  \UseMicrotypeSet[protrusion]{basicmath} 
}{}
\makeatletter
\@ifundefined{KOMAClassName}{
  \IfFileExists{parskip.sty}{%
    \usepackage{parskip}
  }{
    \setlength{\parindent}{0pt}
    \setlength{\parskip}{6pt plus 2pt minus 1pt}}
}{
  \KOMAoptions{parskip=half}}
\makeatother
\usepackage{longtable,booktabs,array}
\usepackage{calc} 
\usepackage{etoolbox}
\makeatletter
\patchcmd\longtable{\par}{\if@noskipsec\mbox{}\fi\par}{}{}
\makeatother
\IfFileExists{footnotehyper.sty}{\usepackage{footnotehyper}}{\usepackage{footnote}}
\makesavenoteenv{longtable}
\usepackage{graphicx}
\makeatletter
\newsavebox\pandoc@box
\newcommand*\pandocbounded[1]{
  \sbox\pandoc@box{#1}%
  \Gscale@div\@tempa{\textheight}{\dimexpr\ht\pandoc@box+\dp\pandoc@box\relax}%
  \Gscale@div\@tempb{\linewidth}{\wd\pandoc@box}%
  \ifdim\@tempb\p@<\@tempa\p@\let\@tempa\@tempb\fi
  \ifdim\@tempa\p@<\p@\scalebox{\@tempa}{\usebox\pandoc@box}%
  \else\usebox{\pandoc@box}%
  \fi%
}
\def\fps@figure{htbp}
\makeatother
\NewDocumentCommand\citeproctext{}{}

\makeatletter
 \let\@cite@ofmt\@firstofone
 \def\@biblabel#1{}
 \def\@cite#1#2{{#1\if@tempswa , #2\fi}}
\makeatother
\newlength{\cslhangindent}
\newlength{\csllabelwidth}
\newenvironment{CSLReferences}[2] 
 {\begin{list}{}{%
  \setlength{\itemindent}{0pt}
  \setlength{\leftmargin}{0pt}
  \setlength{\parsep}{0pt}
  \ifodd #1
   \setlength{\leftmargin}{\cslhangindent}
   \setlength{\itemindent}{-1\cslhangindent}
  \fi
  \setlength{\itemsep}{#2\baselineskip}}}
 {\end{list}}
\usepackage{calc}

\providecommand{\tightlist}{%
  \setlength{\itemsep}{0pt}\setlength{\parskip}{0pt}}
\usepackage{newunicodechar}
\newunicodechar{≈}{\ensuremath{\approx}}
\newunicodechar{≥}{\ensuremath{\geq}}
\newunicodechar{≤}{\ensuremath{\leq}}
\newunicodechar{→}{\ensuremath{\rightarrow}}
\newunicodechar{⇒}{\ensuremath{\Rightarrow}}
\newunicodechar{↔}{\ensuremath{\leftrightarrow}}
\newunicodechar{∧}{\ensuremath{\wedge}}
\newunicodechar{∨}{\ensuremath{\vee}}
\newunicodechar{×}{\ensuremath{\times}}
\newunicodechar{¹}{\ensuremath{^{1}}}
\newunicodechar{²}{\ensuremath{^{2}}}
\newunicodechar{³}{\ensuremath{^{3}}}
\usepackage[htt]{hyphenat}
\usepackage{bookmark}
\IfFileExists{xurl.sty}{\usepackage{xurl}}{} 
\hypersetup{
  colorlinks=true,
  linkcolor={blue},
  filecolor={Maroon},
  citecolor={teal},
  urlcolor={blue},
  pdfcreator={LaTeX via pandoc}}

\author{}
\date{}

\begin{document}

\section{Federated Formal Verification: Cross-Backend Citation,
Cross-Axis Convergence, and AI-Orchestrated Proof Dispatch for
Production
Systems}\label{federated-formal-verification-cross-backend-citation-cross-axis-convergence-and-ai-orchestrated-proof-dispatch-for-production-systems}

\begin{center}
Pierre Falda \\
\texttt{pierre.falda@bullish.com} \\
Bullish
\end{center}

\subsection{Abstract}\label{abstract}

We propose a \emph{federated architecture} for production formal
verification. Rather than forcing all obligations into a single
proof-assistant kernel, the architecture treats a verification campaign
as a polyglot proof system composed of three mechanisms:
\emph{cross-backend citation} discharges a TLA+ obligation by citing an
equivalent theorem in a structurally distinct kernel, with
build-system-level drift-resistance enforced through kernel-level
closure-assertion directives (Coq \texttt{Print\ Assumptions}, Lean 4
\texttt{\#print\ axioms}, and per-kernel analogues --- established
practice in each kernel's community, composed here under a CI-gated
citation primitive); \emph{cross-axis convergence} composes
per-obligation verdicts across independent verifiers into operational
kernel-agreement gates (the agreement is build-system-enforced rather
than a logical soundness theorem; the residual
correspondence-certificate question is named as an open problem in §6.3
and partially checked empirically in §5.5); \emph{AI-orchestrated
mass-parallel dispatch} fans out autonomous specialist agents in waves
with disjoint file-touch zones. Verification authority resides at the
kernel-directive and CI-gated-citation layer; the AI layer is untrusted
proof-search labour inside a trusted CI envelope. We validate the
architecture on two production subsystems of the Mercury
high-frequency-trading platform: a Raft consensus subsystem with full
algorithmic scope (joint consensus, leadership transfer, log compaction,
linearizable client reads, dynamic reconfiguration) and a
financial-arithmetic invariant layer (balance accounting,
automated-market-maker curve invariants, isolated-margin, lock-tracking
settlement). The validation campaign reduced a 26-axiom Raft census to
zero in 17 active hours of single-session wallclock, an empirical ≈ 60×
per-axiom reduction against the team's prior Path-A.2 TLAPS
ghost-composition baseline at intra-shape scope (§5.2.2; classified
\emph{author-prior} under the §5.2 comparator-hygiene template); a
within-method comparator (§5.2.1; \emph{author-generated}) and a
cross-precedent IronFleet comparator (§5.2.3;
\emph{independent-published} under three explicit scope qualifiers ---
cross-protocol, cross-scope, cross-rigour) report consistent
direction-of-effect. The ratios are a property of 2026 AI-agent compute
economics and should be expected to evolve. The campaign also brought
five numeric canonicals to 6-axis chorus discharge, and surfaced four
real production bugs --- including a perp-funding 10¹² over-charge
structurally invisible to single-backend specification campaigns.
Compared to verified-Raft precedents (IronFleet, Verdi-Raft,
MongoRaftReconfig), the architecture delivers broader coverage on
algorithmic-scope dimensions (joint consensus, leadership transfer, log
compaction, linearizable client reads, dynamic reconfiguration) at
one-to-two orders of magnitude lower per-axiom wallclock under the same
three scope qualifiers as §5.2.3 (cross-protocol, cross-scope,
cross-rigour), against a trusted base that is larger than what the
verified-extraction precedents accept (the present method does not claim
verified extraction). The principal open problem is the mechanisation of
\emph{correspondence certificates} --- equivalence between a TLA+
obligation and its cited external theorem, currently checked empirically
by the cross-axis matrix but not yet mechanically proved.

\begingroup\raggedright \textbf{Keywords}: formal verification, distributed consensus,
cross-backend proof citation, AI-orchestrated proof automation, TLA+,
Coq, Lean 4, Apalache, multi-prover convergence, financial-system
verification, mechanical theorem proving.\par\endgroup

\subsection{1. Introduction}\label{introduction}

\subsubsection{1.1 The cost-of-rigour
frontier}\label{the-cost-of-rigour-frontier}

Mechanical verification of production software systems faces a
persistent cost-of-rigour frontier defined by three structural
constraints.

\emph{Backend monoculture}. A verification campaign typically rests on a
single proof-assistant kernel, so a kernel soundness bug, an automation
regression, or a stdlib axiom shift becomes a single point of failure
for the entire artefact. Every prior mechanical proof of a Raft-family
safety property uses exactly one verifier (TLAPS for Ongaro (Ongaro
2014) and MongoRaftReconfig (Schultz et al. 2022); Coq for Verdi-Raft
(Wilcox et al. 2015; Woos et al. 2016) and Velisarios (Rahli et al.
2018); Dafny + Z3 for IronFleet (Hawblitzel et al. 2015); Ivy + Z3 for
the Padon and Scimitar lines (Padon et al. 2017, 2024; Schultz et al.
2024); Verus for Anvil (Sun et al. 2024; Lattuada et al. 2024)).

\emph{Excluded algorithmic scope}. Published verified protocols
routinely omit features the production deployment depends on.
Verdi-Raft's PLDI 2015 paper (Wilcox et al. 2015) explicitly excludes
reconfiguration, log compaction, and linearizable client reads, all of
which any production Raft library must implement; the CPP 2016 follow-up
(Woos et al. 2016) formalises additional Raft extensions but does not
close the production gap. The TigerBeetle Viewstamped Replication
challenge excluded reconfiguration and snapshot transfer; etcd's TLA+
specification covers a subset of state transitions; CCF's authors note
their consensus is ``based on an unproven algorithm'' (Howard et al.
2025).

\emph{Per-axiom serialisation}. The dominant cost of a
multi-thousand-step proof is iteration on a single specialist's working
memory, scaling linearly with proof complexity and rate-limiting the
multi-person-year campaigns reported in the precedents (IronFleet 3.7
person-years for IronRSL; Verdi-Raft ≈ 2 person-years community estimate
for the original Raft proof; MongoRaftReconfig ≈ 4 person-months for
Raft reconfiguration safety).

Each constraint is independently addressable in principle, but published
precedents address subsets only. The literature trades each constraint
against the others. The common thread across the precedent set is that
backend monoculture, scope exclusion, and serialisation are
\emph{coupled}: relaxing any one is observed to require infrastructure
that the precedent's authors did not invest in for the published
artefact.

\subsubsection{1.2 Why this matters}\label{why-this-matters}

The three constraints have braked the diffusion of formal verification
into industrial practice for thirty years. Hoare's Verified Software
Grand Challenge (Hoare 2003) identified exactly this cost frontier as
the rate-limiting step in industrial adoption and framed it as a problem
requiring sustained community effort. The present architecture attacks
the frontier by composition of orthogonal mechanisms --- cross-backend
citation, cross-axis convergence, and AI-orchestrated dispatch ---
rather than by an advance in any single kernel. Practitioners who
attempt formal verification at production scale routinely abandon the
effort when the projected wallclock exceeds the project's tolerance ---
the reason the production Raft library landscape (etcd, TiKV,
CockroachDB, ScyllaDB, jgroups-raft, Atomix Copycat, Real Logic's Aeron
Cluster, TigerBeetle's VSR) ships with no published mechanical
verification of any specific implementation, despite the well-documented
failure modes that Jepsen-style fault-injection testing surfaces in
those libraries.

The opportunity is large because the cost-of-rigour frontier is the
rate-limiting step in industrial adoption rather than a fundamental
property of mechanical verification. An architecture that compresses the
projected wallclock by one or two orders of magnitude on the constrained
dimension, \emph{without weakening the soundness claim}, opens the
frontier to a substantially broader class of production systems than the
precedent posture admits.

The stakes are concrete: the validation campaign reported in section 4.3
caught a 10¹² over-charge of accrued funding in a perp-funding
settlement path --- a silent multiplicative scale mismatch in a
financial-arithmetic layer that is structurally invisible to the
spec-driven verification campaigns of the precedent literature, none of
which model a financial-arithmetic layer with mechanically-checked
numeric invariants.

\subsubsection{1.3 Contribution: a federated architecture for production
formal
verification}\label{contribution-a-federated-architecture-for-production-formal-verification}

This paper proposes a \emph{federated architecture} for production
formal verification. Instead of forcing all proof obligations into a
single proof assistant or model checker, we treat a production
verification campaign as a polyglot proof system: TLA+ expresses
system-level obligations, specialist backends discharge obligations
suited to their proof shape, CI-enforced citation gates preserve
drift-resistance across the heterogeneous trust base, and AI agents
coordinate parallel proof discovery without entering the trusted
computing base. The result is not a replacement for proof kernels but an
\emph{orchestration layer around them}. We validate the architecture on
two production subsystems of the Mercury trading platform: a full-scope
Raft consensus subsystem and a financial-arithmetic invariant layer
(section 4).

The architecture has three composable mechanisms and one methodological
distinction.

\emph{Cross-backend citation} (section 3.1) is a build-system-gated
primitive that discharges a TLA+ proof obligation by citing an
equivalent theorem in an external proof system --- Coq, Lean 4, Why3,
Apalache, hand-encoded SMT-LIB v2 dispatched to standalone Z3, or CBMC
--- with the build system re-checking the cited theorem under its own
kernel on every commit and failing the build loudly on any drift. The
primitive eliminates backend monoculture per-obligation: the
obligation's discharge transitively depends on the cited kernel, which
is structurally distinct from the host kernel (TLAPS). The semantic
equivalence between the TLA+ obligation and the cited external theorem
is a \emph{correspondence certificate} (section 3.1.4) --- a first-class
component of the architecture's trust base, currently human-asserted,
that the cross-axis convergence matrix partially checks empirically
(section 5.5) but does not yet mechanically prove.

\emph{Cross-axis convergence} (section 3.2) records per-obligation
verdicts across N independent verifiers and composes them into dual- and
triple-kernel agreement gates that fail loud if any one axis regresses.
The composition is operational rather than documentary: the build system
enforces the agreement, not an out-of-band attestation. The epistemic
value scales with paradigm disjointness --- deductive proof versus
bounded symbolic model checking versus probabilistic model checking
carry stronger orthogonality than within-paradigm variants (Coq vs Lean,
which share CIC ancestry).

\emph{AI-orchestrated mass-parallel dispatch} (section 3.3) compresses
the campaign-level wallclock by fanning out N autonomous specialist
agents --- one per verifier or one per obligation --- coordinated by a
single survey-and-route agent that infers the appropriate backend per
obligation and dispatches in waves with disjoint file-touch zones. The
wallclock compression is empirical: section 5.2 reports a measured ≈ 60×
per-axiom reduction against the team's prior Path-A.2 TLAPS
ghost-composition baseline (§5.2.2), with a within-method comparator
(§5.2.1) and a cross-precedent IronFleet comparator (§5.2.3)
corroborating direction-of-effect under named scope qualifiers.
Critically, the AI layer is \emph{untrusted proof-search labour} inside
a \emph{trusted CI envelope} (section 3.3): agents propose, attempt, and
report negatives; verification authority resides with kernel-level
closure-assertion directives --- the established
\texttt{Print\ Assumptions} directive in Coq (shipped since Coq 8.5 in
2016, used as an end-of-development audit gate in
Common-Criteria-adjacent verified-software submissions), the analogous
\texttt{\#print\ axioms} in Lean 4, and the corresponding prover-cascade
and counterexample-exhaustion checks for each wired backend --- composed
with the CI-gated citation primitive of section 3.1. The composition is
the novel contribution; the per-kernel directives it composes are
established practice in their respective communities. This separation is
load-bearing --- it allows agent parallelism to scale without expanding
the trusted computing base.

\emph{The algorithmic-correctness-versus-annotation-depth distinction}
(section 3.4) separates the load-bearing soundness chain (TLA+
specification → TLAPS proof → annotation-processor-generated runtime
invariant verifier) from the additional epistemic redundancy of the
cross-axis chorus. The distinction supports honest epistemic accounting:
a regression in an annotation-layer axis port does not invalidate the
algorithmic claim because the algorithmic chain remains intact.

\begin{center}\includegraphics[width=\textwidth,height=0.85\textheight,keepaspectratio]{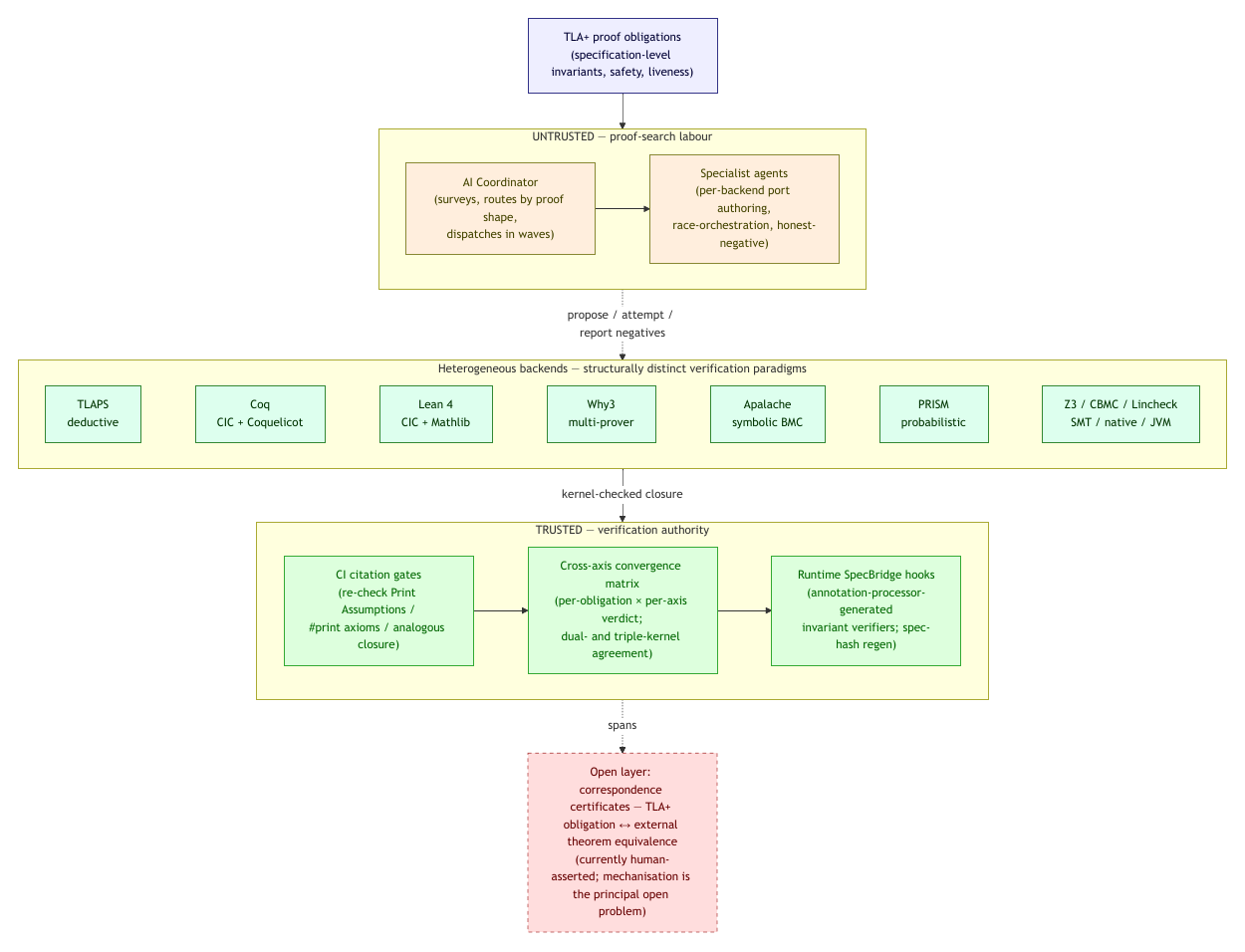}\end{center}

\emph{Figure 1. Trust-hierarchy overview of the federated verification
architecture. TLA+ proof obligations enter at the top. The AI layer
(UNTRUSTED --- proof-search labour) comprises a coordinator that surveys
gaps and routes by proof shape and N specialist agents that author
per-backend ports, race-orchestrate, and report negatives honestly.
Specialists dispatch to a heterogeneous backend roster spanning
deductive (TLAPS, Coq, Lean 4, Why3), bounded (Apalache), probabilistic
(PRISM), SMT (standalone Z3), and production-language (CBMC, Lincheck)
verifiers. The TRUSTED layer comprises the CI citation gates (which
re-check \texttt{Print\ Assumptions} / \texttt{\#print\ axioms} /
analogous closure on every commit), the cross-axis convergence matrix
(per-obligation × per-axis verdict, composed into dual- and
triple-kernel agreement gates), and the runtime SpecBridge hooks
(annotation-processor-generated invariant verifiers with spec-hash-keyed
regeneration that closes the implementation-versus-specification gap at
compile time). The open dashed layer is the }correspondence certificate*
--- the equivalence between a TLA+ obligation and the external theorem
cited to discharge it. Currently human-asserted; mechanisation is the
principal open problem the architecture surfaces (sections 3.1.4, 6.3,
6.4).*

\subsubsection{1.4 Why this is new}\label{why-this-is-new}

The three constraints have been individually relaxed in recent
precedents. Bythos (Zhao et al. 2024) embeds TLA semantics into Coq for
Byzantine-protocol composition, addressing the backend boundary at the
language-design layer. CCF Smart Casual Verification (Howard et al.
2025) integrates TLC model-checking continuously into CI to address
scope-exclusion via fast iteration on a model-checked artefact. Anvil
(Sun et al. 2024) verifies Kubernetes controller liveness in Verus with
TLA-embedding, demonstrating that production-scale liveness is tractable
on a verifier-with-embedding stack.

The common limitation across these recent precedents is that each
pursues a single axis of relaxation. None addresses backend monoculture
by allowing arbitrary external citations from a host proof system; none
addresses serialisation by parallelising across an autonomous agent
fleet; none combines the relaxations into a method that closes complete
classes of obligations across heterogeneous proof shapes within a single
engineering platform.

The advance the present method offers is \emph{compositionality} of the
three relaxations. The cross-backend citation primitive, the cross-axis
convergence matrix, and the AI-orchestrated dispatch protocol are
independently useful, but their composition is what makes a 26-axiom
Raft census closeable in seventeen single-session hours rather than
five-to-seven person-months. The same composition also closes a
financial-arithmetic invariant layer with heterogeneous proof shapes ---
demonstrating that the method generalises across proof-shape classes
within a single engineering substrate.

\subsubsection{1.5 Validation strategy}\label{validation-strategy}

We validate the method on two subsystems of the Mercury
high-frequency-trading platform --- a production-deployed
distributed-consensus subsystem and a production-deployed
financial-arithmetic invariant layer. The validation has two purposes.
First, it confirms that the method scales to a real production system
with all the engineering constraints of one (large existing codebase,
performance constraints, deployment pressure). Second, the
cross-subsystem application --- same toolchain, same dispatch protocol,
same citation primitive on two structurally distinct proof-shape classes
--- establishes that the method is a method rather than a one-off
tactical alignment with one subsystem's incidental properties.

Mercury's Raft is a production-deployed implementation of Diego Ongaro's
Raft (Ongaro 2014; Ongaro and Ousterhout 2014) with joint consensus,
leadership transfer, log compaction, linearizable client reads, and
dynamic reconfiguration. Mercury's Themis matching engine implements
balance accounting, automated-market-maker liquidity-pool curve
invariants, an isolated-margin product family, and lock-tracking
settlement. Both subsystems run inside the same JVM process in the
production deployment.

\subsubsection{1.6 Paper organization}\label{paper-organization}

Section 2 surveys related work --- academic verified-consensus
precedents, big-tech industrial TLA+ practice, the production Raft
library landscape, the broader verified-systems literature,
cross-backend verification, and AI-assisted formal verification --- and
identifies six recurring compromises against which the architecture is
positioned. Section 3 presents the architecture's three primary
mechanisms plus the algorithmic-versus-annotation distinction. Section 4
reports the cross-subsystem validation. Section 5 reports performance:
the wallclock benchmark against single-specialist and IronFleet-baseline
comparators, the cross-axis convergence matrix empirics, the
production-bug-finding empirics, and honest negative results. Section 6
discusses limitations, threats to validity, and the open problem of
mechanising correspondence certificates. Section 7 concludes.
Acknowledgements appear in section 8; references in section 9. Appendix
A bundles per-claim verification recipes that produce every numerical
and architectural claim in this paper from the published source
repository.

\subsection{2. Related work}\label{related-work}

A method-paper claim is meaningful only against a profiled prior-art
baseline. This section discharges that bookkeeping. Section 2.1 surveys
mechanical verification of Raft and adjacent consensus protocols --- the
prior art most directly comparable to the validation subsystem. Section
2.2 surveys the broader verified-systems literature that informs the
method-design choices. Section 2.3 surveys cross-backend verification,
the closest published precedent for the citation primitive. Section 2.4
surveys AI-assisted formal verification, the closest published precedent
for the dispatch protocol. Section 2.5 catalogues six recurring
compromises observable across the precedent set, against which the
architecture is positioned.

\subsubsection{2.1 Mechanical verification of Raft and adjacent
consensus}\label{mechanical-verification-of-raft-and-adjacent-consensus}

The published mechanical-verification literature for Raft and adjacent
consensus protocols spans approximately fifteen years of academic and
industrial activity. Table 1 enumerates the precedents we survey,
recording for each the verifier(s) used, the proven scope, whether
liveness was mechanically discharged, and whether the verified artefact
reached production deployment. The table is deliberately broad: it
includes every published mechanical proof of Raft we can identify, every
recent industrial spec-and-model-checking effort touching Raft-family
protocols, the Byzantine-fault-tolerant precedents that share
methodological DNA with the cross-axis approach, and the production Raft
library landscape that ships without published mechanical verification.

{\def\LTcaptype{none} 
\begingroup\tiny\setlength{\tabcolsep}{2pt}\renewcommand{\arraystretch}{0.92}\begin{longtable}[]{@{}
  >{\raggedright\arraybackslash}p{(\linewidth - 10\tabcolsep) * \real{0.1071}}
  >{\raggedright\arraybackslash}p{(\linewidth - 10\tabcolsep) * \real{0.1607}}
  >{\raggedright\arraybackslash}p{(\linewidth - 10\tabcolsep) * \real{0.2143}}
  >{\raggedright\arraybackslash}p{(\linewidth - 10\tabcolsep) * \real{0.1250}}
  >{\raggedright\arraybackslash}p{(\linewidth - 10\tabcolsep) * \real{0.1786}}
  >{\raggedright\arraybackslash}p{(\linewidth - 10\tabcolsep) * \real{0.2143}}@{}}
\toprule\noalign{}
\begin{minipage}[b]{\linewidth}\raggedright
Year
\end{minipage} & \begin{minipage}[b]{\linewidth}\raggedright
Project
\end{minipage} & \begin{minipage}[b]{\linewidth}\raggedright
Backend(s)
\end{minipage} & \begin{minipage}[b]{\linewidth}\raggedright
Scope
\end{minipage} & \begin{minipage}[b]{\linewidth}\raggedright
Liveness
\end{minipage} & \begin{minipage}[b]{\linewidth}\raggedright
Production
\end{minipage} \\
\midrule\noalign{}
\endhead
\bottomrule\noalign{}
\endlastfoot
2014 & Ongaro PhD \texttt{raft.tla} (Ongaro 2014) & TLA+ + TLAPS &
LogCompleteness mechanically proved against admitted invariants & No &
Specification only \\
2015 & Verdi-Raft (Wilcox et al. 2015; Woos et al. 2016) & Coq & State
Machine Safety + linearizability; \textasciitilde2 person-years
(community estimate) & No & Research artefact (\texttt{vard}) \\
2015 & IronFleet / IronRSL (Hawblitzel et al. 2015, 2017) & Dafny + Z3 &
Multi-Paxos safety + liveness; verified extraction to executable C\#;
\textasciitilde3.7 person-years & Yes & Microsoft research artefact \\
2015 & AWS use of TLA+ (Newcombe et al. 2015) & TLA+ + TLC & DynamoDB,
S3, EBS --- model-checked, not theorem-proved & No & TLA+ specs guided
design \\
2016 & Chand--Liu--Stoller Multi-Paxos (Chand et al. 2016) & TLA+ +
TLAPS & Multi-Paxos full safety & No & Research \\
2017 & Padon--Losa--Sagiv--Shoham Paxos-EPR (Padon et al. 2017) & Ivy +
Z3 & Safety of Paxos, Multi-Paxos, Vertical Paxos, Fast Paxos, Flexible
Paxos, Stoppable Paxos & No & Research \\
2018 & Liveness-to-safety reduction (Padon et al. 2018) & Ivy + Z3 &
First mechanised liveness for Multi-Paxos via reduction & Yes &
Research \\
2018 & Disel (Sergey et al. 2018) & Coq + Hoare Type Theory & TPC, query
layer & No & Research framework \\
2018 & Velisarios (Rahli et al. 2018) & Coq + logic-of-events & First
mechanically-checked PBFT safety & No & Research \\
2020 & Aneris (Krogh-Jespersen et al. 2020) & Coq + Iris & TPC +
replicated logging & No & Research \\
2022 & MongoRecfg (Schultz et al. 2022) & TLA+ + TLAPS & LeaderCompl. +
StateMachineSafety & No & Ships in MongoDB \\
2022 & Azure Cosmos DB TLA+ (Hackett et al. 2022) & TLA+ + TLC & Cosmos
DB consistency model & No & Cosmos DB ships \\
2022 & Oracle Labs \texttt{bft-consensus-agda} (Carr et al. 2022) & Agda
& HotStuff, LibraBFT, DiemBFT safety & No & Research; archived \\
2024 & Verus (Lattuada et al. 2024) & Verus (Rust → Z3) & Foundation for
Rust systems verification & n/a & Used at Microsoft, Amazon \\
2024 & Anvil (Sun et al. 2024) & Verus + TLA-embedding & Liveness of
Kubernetes controllers & Yes & Verified controllers \\
2024 & etcd-raft TLA+ + trace validation & TLA+ + TLC + trace validation
& Subset of state transitions & No & etcd ships \\
2024 & mypyvy / Scimitar (Padon et al. 2024; Schultz et al. 2024) & Ivy
+ Z3 & Inductive AbstractRaft + AsyncRaft safety & No & Research \\
2024 & Bythos (Zhao et al. 2024) & Coq + TLA-in-Coq & Byzantine
composition safety + liveness & Yes & Research \\
2025 & CCF Smart Casual (Howard et al. 2025) & TLA+ + TLC + sim + trace
& Modified Raft model in CI; 6 bugs caught pre-prod & Yes (MC) & Azure
Conf. Ledger \\
2025 & DAG-based BFT verification (NFM 2025) & TLA+ + TLAPS & DAG-Rider,
Cordial Miners, Hashgraph, BullShark, Aleph & No & Research \\
2026 & LeaseGuard ({Davis et al.} 2026) & TLA+ + TLC + Python sim +
LogCabin C++ & Raft leader-lease, Read-Your-Writes & No & LogCabin
(research) \\
\end{longtable}\endgroup
}

\emph{Table 1. Mechanical-verification precedents for Raft and adjacent
consensus.}

The production Raft library landscape carries an additional class of
relevant prior art that does not fit Table 1's inclusion criteria but
bears mention. The widely-used production Raft libraries --- etcd's
\texttt{etcd-io/raft} (powering Kubernetes, HashiCorp Consul, Apache
Kafka KRaft), TiKV/TiDB's Rust port of etcd-raft, CockroachDB's Raft
layer, YugabyteDB's Raft with leader-lease enhancements, ScyllaDB's Raft
(used for schema and topology coordination), Atomix Copycat, Real
Logic's Aeron Cluster (used in financial trading systems), and
TigerBeetle (Viewstamped Replication rather than Raft) --- all ship
without their own published mechanical verification of their specific
implementations. etcd added a partial TLA+ specification and a
trace-validation tool in 2024; TiKV documents that its membership-change
protocol diverges from the paper for ``simple implementation''; the
others rely primarily on Jepsen-style fault-injection testing.

We could not identify published mechanical formal verification of
Google's Spanner or Chubby; Google has not published proofs of their
internal consensus implementations (the underlying Multi-Paxos has been
verified independently by Chand--Liu--Stoller as catalogued in Table 1).
We could not identify published mechanical proof of Meta's HotStuff or
DiemBFT consensus implementations; the closest proxy is the Oracle Labs
work catalogued above. We identified no published TLA+ or
mechanical-verification consensus papers from Apple. IBM Research has
published verification of customisable Hyperledger Fabric consensus
rules but not a Raft proof.

\subsubsection{2.2 The broader verified-systems
literature}\label{the-broader-verified-systems-literature}

The verified-systems tradition extends beyond consensus to verified
compilers (CompCert (Leroy 2006)), verified operating-system kernels
(seL4 (Klein et al. 2009)), verified cryptographic libraries (F* +
Project Everest (Swamy et al. 2016; Bhargavan et al. 2017)), verified
distributed key-value stores (Chapar, Cure (Lesani et al. 2016;
Akkoorath et al. 2016)), and verified Rust-based systems software (Verus
(Lattuada et al. 2024); Anvil (Sun et al. 2024)). The pattern across
this literature is that a single backend (Coq for CompCert and seL4; F*
for Project Everest; Verus for Anvil) is selected per project and the
entire artefact rests epistemically on that backend's kernel. None of
the precedents in this tradition cross-checks the same theorem across
multiple structurally distinct kernels.

The matching-engine financial-arithmetic verification literature is
comparatively thin. Most exchanges treat the matching engine as a
hand-coded high-performance system gated by integration tests, fuzz
suites, and property-based tests rather than mechanical proofs. We could
not identify published mechanical formal verification of a production
matching engine's arithmetic invariant layer; the closest adjacent
comparators (TigerBeetle's deterministic simulation testing) operate at
a methodologically distinct verification class (trace-validation rather
than theorem-proving).

\subsubsection{2.3 Cross-backend
verification}\label{cross-backend-verification}

Cross-backend verification has appeared informally across F* + Z3 (where
F* discharges Hoare triples to Z3 and accepts Z3's verdict as proof),
Lean Mathlib citing Coq results, and Isabelle's Sledgehammer dispatching
to Z3, SPASS, and CVC. We classify the published precedents along four
classes of cross-backend integration; the citation primitive of section
3.1 occupies a fifth position structurally distinct from all four.

A first class is \emph{whole-language semantic embedding}, exemplified
by Bythos (Zhao et al. 2024), which embeds TLA's temporal operators into
Coq via \texttt{CoFixpoint} constructs and proves the embedded-TLA
propositions in Coq directly. Bythos and the present method are
architecturally distinct: Bythos \emph{embeds} TLA semantics into Coq
and proves the embedded-TLA propositions in Coq's native language,
whereas the citation primitive of section 3.1 leaves TLA semantics in
TLAPS, exhibits a sister theorem in the external kernel's native
vocabulary, and uses the build system as the meta-system that ensures
the sister theorem closes and remains cited. Section 3.1.4 develops this
comparison further.

A second class is the \emph{whole-tool trust delegation} pattern (F* +
Z3 discharges Hoare triples to Z3 internally, accepting Z3's
\texttt{unsat} verdict as proof). The trust regime there is ``Z3 is
sound on the SMT-LIB fragment F* generates.'' The citation primitive's
trust regime is narrower: the cited external theorem is a specific named
theorem whose closure is asserted by the kernel's own
\texttt{Print\ Assumptions}-class directive, and the build-system gate
verifies that specific assertion.

A third class is \emph{de novo embedding of one logic into another} ---
for instance, Sledgehammer's translation of higher-order Isabelle goals
into first-order SMT, or Coq's \texttt{coq-of-ocaml} translation of
OCaml programs into Coq. These translations require an embedding
correctness argument that the method's per-obligation citation primitive
does not require.

A fourth class is \emph{build-system-integrated single-kernel
verification}, exemplified by LiquidHaskell (Vazou et al. 2014) and
especially its GHC-plugin reframing (Di Napoli et al. 2020), in which
refinement-type checking is wired into the host language compiler and a
refinement violation aborts the build with a non-zero exit code. The
integration mechanism is single-source-language (Haskell) and
single-kernel (the Z3 verdict on the SMT encoding); the
build-system-gated discipline (compile-time, fail-loud) is the direct
precedent for the citation primitive's Gradle gate of section 3.1.2. The
present method extends from the single-source-language single-kernel
posture to the multi-source-language multi-kernel composition described
in section 3.1.

\subsubsection{2.4 AI-assisted formal
verification}\label{ai-assisted-formal-verification}

AI-assisted formal verification has a sparse precedent literature in
2024--2026, but the precedent set is concentrated in a single regime ---
\emph{LLM-driven proof search inside one kernel}. The DeepMind
AlphaGeometry line (Trinh et al. 2024) uses an LLM as a tactic-search
guide over a single custom geometry proof kernel (with the AlphaProof
successor extending the same regime to Lean). Lean Copilot (Song et al.
2025) integrates LLM-driven tactic suggestion and premise selection into
the Lean 4 interactive proof workflow. Sledgehammer (Paulson and
Blanchette 2010), the longest-running precedent of this kind, dispatches
Isabelle goals across an automation cascade of external first-order
provers (E, SPASS, CVC, Z3) and translates the discovered proof back
into Isabelle's kernel. Each of these is \emph{single-host-kernel by
construction}: the LLM (or premise-selector, or automation cascade)
lives inside one proof system and helps that system close its own
obligations faster.

A second precedent set sits one rung lower in the trust stack ---
LLM-agent orchestration patterns that compose multi-step reasoning
without delegating verification authority. Plan-then-execute patterns
(The LangChain Team 2023; Wang et al. 2023) decompose a task into an
explicit plan-step followed by execution-steps (Plan-and-Solve adds an
explicit ``plan first, then solve'' chain-of-thought instruction at the
prompt layer; LangChain wraps the same decomposition in an executor
harness). Self-critique patterns --- Reflexion (Shinn et al. 2023) and
Self-Refine ({Madaan et al.} 2023) --- establish the discipline that an
agent must surface what it could not accomplish, feeding that admission
into the next iteration (Reflexion adds an explicit verbal-reinforcement
loop on negative outcomes; Self-Refine iterates within a single task on
self-generated critique). Multi-agent debate (Du et al. 2023) composes
independent agent verdicts into agreement-via-disagreement. Process
reward models (Lightman et al. 2023) reward per-step reasoning rather
than only final outcome. The dispatch architecture of section 3.3
occupies a structurally distinct position from each of these: it lifts
the honest-negative discipline (Reflexion-family verbal-reflection on
negatives) into a coordinator-mediated \emph{cross-axis re-routing}
primitive (extending plan-then-execute decomposition with a per-step
structurally-typed backend-roster re-dispatch rather than within-tool
retry), uses the per-step verdict as a process reward to route the next
dispatch (extending the PRM intuition into an orchestration loop), and
substitutes kernel-verified PROVED for the agreement signal that
multi-agent debate must approximate via cross-model consensus. The
verdicts that drive routing decisions are kernel-emitted closure
assertions, not LLM self-assessments --- this is the load-bearing
departure from the precedent set.

Concretely, the dispatch architecture is not a single-kernel
tactic-search aid (as the proof-search-side precedent set is); it is a
\emph{coordinator-and-specialist topology for proof campaigns} in which
specialist agents author per-backend ports across the citation
primitive's heterogeneous kernel roster (TLAPS / Coq / Lean 4 / Why3 /
Apalache / PRISM / standalone Z3 / CBMC / Lincheck) and a coordinator
routes obligations by static signal. The agents do not enter the trusted
computing base --- verification authority remains with kernel-level
\texttt{Print\ Assumptions}-class directives and the build-system
citation gates (section 3.3 design principle). The two design-space
differences from the single-kernel proof-search precedent set are
therefore (i) \emph{cross-backend campaign orchestration} rather than
single-backend tactic search, and (ii) \emph{honest-negative discipline}
(section 3.3.3) as the architectural choice that makes autonomous proof
search epistemically sound --- every dispatched specialist must surface
what it could not close, which the coordinator then cross-routes to a
structurally appropriate alternative. The dispatch protocol's empirical
impact reported in section 5.2 should be expected to evolve as the
underlying agent economics evolve; the \emph{architectural} contribution
(the coordinator-and-specialist topology with honest-negative reporting
and trust-base separation) is the durable claim.

\subsubsection{2.5 Six recurring
compromises}\label{six-recurring-compromises}

Six recurring compromises stand out across the precedent set. We
catalogue them here and return to the architecture's positioning against
each in section 3.

\emph{Single-backend monoculture} is universal across the mechanical
Raft and Raft-family proofs in Table 1. None cross-checks the same
theorem across multiple independent kernels with disjoint trusted bases.

\emph{Liveness deferral} is the rule rather than the exception. Only
IronFleet, the Padon et al.~2018 liveness-to-safety reduction (Padon et
al. 2018), Anvil, Bythos, and CCF (model-checked) mechanically discharge
liveness. The canonical Raft proofs (Ongaro, Verdi-Raft,
MongoRaftReconfig, Scimitar) are safety-only.

\emph{Scope exclusions} are pervasive. Verdi-Raft excludes
reconfiguration and log compaction; the TigerBeetle VSR challenge
excluded reconfiguration and snapshot transfer; etcd's TLA+
specification covers a subset of state transitions; CCF's authors
explicitly note that their consensus is ``based on an unproven
algorithm.''

\emph{Specification-versus-proof conflation in industrial reporting} is
widespread. The AWS, Azure Cosmos DB, MongoDB \texttt{RaftMongo.tla},
etcd TLA+, and CCF efforts are TLC model-checking and trace-validation
on bounded state spaces --- not deductive theorem-proving --- and
characterising them as ``verified'' without that qualification
overstates the epistemic status of the result.

\emph{Implementation-versus-specification gap} is rarely closed. Even
where a mechanical proof exists, the proven Coq, Dafny, or TLA+ artefact
is rarely the production binary. Trace-validation projects (etcd, CCF,
MongoDB conformance-checking) acknowledge this gap explicitly and
address it post hoc rather than at proof construction time.

\emph{Single-operator serialisation of proof campaigns} is the cost
driver behind the multi-person-year wallclock totals reported across the
precedent set. IronFleet's 3.7 person-years, Verdi-Raft's ≈ 2
person-years, MongoRaftReconfig's ≈ 4 person-months, and CompCert's even
longer cumulative effort all reflect a \emph{single specialist iterating
on one obligation at a time}, with throughput rate-limited by one
expert's working memory. The compromise is structural rather than
methodological --- every precedent in Table 1 adopts it implicitly
because no precedent invests in a multi-specialist coordination layer
that preserves the soundness claim. Section 3.3 attacks this compromise
directly via the AI-orchestrated dispatch protocol.

The architecture we describe in section 3 is positioned against each of
these compromises by a structural mechanism rather than by additional
engineering effort within a single-backend posture.

\subsection{3. The architecture}\label{the-architecture}

The federated verification architecture has three primary mechanisms and
a fourth methodological distinction. Section 3.1 describes the
cross-backend citation primitive. Section 3.2 describes the cross-axis
convergence matrix. Section 3.3 describes the AI-orchestrated
mass-parallel dispatch protocol. Section 3.4 introduces the
algorithmic-correctness-versus-annotation-depth distinction that
separates the load-bearing soundness chain from the annotation-layer
chorus.

\subsubsection{3.1 Cross-backend citation
primitive}\label{cross-backend-citation-primitive}

The cross-backend citation primitive discharges a TLA+ proof obligation
by citing an equivalent theorem in an external proof system, with the
build system re-checking the cited theorem on every commit and failing
loud on any drift. This section describes the primitive's design
(3.1.1), the build-system mechanism (3.1.2), the dual- and
triple-citation gates that compose the primitive into kernel-agreement
gates (3.1.3), and the architectural distinction between the primitive
and the embedding approach Bythos takes for the same boundary (3.1.4).

\paragraph{3.1.1 Design}\label{design}

A TLA+ proof obligation can be discharged in three ways. The first is a
direct TLAPS deductive proof, in which the theorem's \texttt{PROOF}
block is checked by tlapm's hierarchical proof system and dispatched
through the standard Zenon → Isabelle → SMT-cascade chain (Chaudhuri et
al. 2010; Cousineau et al. 2012). The second is the cross-backend
citation, in which the theorem's \texttt{PROOF} block is replaced by
\texttt{PROOF\ OMITTED} annotated with a structured \texttt{CITED:} tag
naming an external proof-system theorem whose closure discharges the
obligation. The third is a sister-spec deduction, in which the theorem
is annotated
\texttt{PROOF\ OMITTED\ \textbackslash{}*\ DEDUCTIVE-DISCHARGE-IN-SISTER:}
to indicate that the obligation is deductively discharged in a companion
specification module.

The cross-backend citation is the primitive of interest. The structured
\texttt{CITED:} tag carries three pieces of information: the path of the
external proof file, a \texttt{\#}-separated theorem identifier within
that file, and (for dual- and triple-citation rows)
\texttt{\textbar{}\textbar{}}-separated alternate-kernel references for
each independent axis the obligation is cited against. The
build-system-level mechanism that wires the citation enforces three
properties: the cited external theorem must exist in the cited file; the
external proof file must check cleanly under its own kernel; and the
kernel's \texttt{Print\ Assumptions} (or analogous) output for the cited
theorem must report closure under the global trusted base with no
\texttt{Admitted}, \texttt{Axiom}, or \texttt{Parameter} declarations
beyond a narrow whitelist of the kernel's standard library. This use of
\texttt{Print\ Assumptions} extends the directive's established role as
an end-of-development audit gate (e.g., \texttt{Print\ Assumptions}
reports submitted with high-assurance verified-software evaluation
packages) by making the assertion a CI-gated invariant rather than a
one-time check, and by composing assertions across structurally distinct
kernels rather than within a single one.

\paragraph{3.1.2 Build-system mechanism}\label{build-system-mechanism}

The primitive is realised as a Gradle task that wires three checks into
the build graph at the layer that holds the TLAPS proof. On every build
of the parent TLAPS check task, the wired citation task fires three
sub-checks. First, the cited TLA+ proof file is grepped for the literal
pattern \texttt{\^{}THEOREM\ \textless{}theoremName\textgreater{}}
followed by a \texttt{CITED:} annotation containing the cited external
file and theorem name; an absent or drift-broken annotation throws a
build exception with a precise diagnostic. Second, the cited external
proof file's verification log is parsed for the
\texttt{Print\ Assumptions} (or \texttt{\#print\ axioms}, or analogous)
output for the cited theorem; an absent or impure closure throws. Third,
defensively, the count of closure-assertion directives in the external
proof file is compared against the count of closure-confirmation lines
in the verification log to ensure each asserted closure has a
corresponding kernel-confirmed pass.

The combination of these three checks makes the primitive
\emph{drift-resistant by construction}. There is no operator-driven
attestation step, no manual citation register, no out-of-band log of
``the external proof was clean as of date X.'' The attestation is the
build itself, and the build is run on every commit through the unit-test
layer, which gates every pull request.

The Gradle DSL that wires the citation primitive
(\texttt{tla.registerCoqExternalCitation(...)},
\texttt{tla.registerLeanExternalCitation(...)}, and the dual- and
triple-axis siblings) functions as a lightweight \emph{meta-logical
framework} in the sense familiar from the logical-frameworks literature:
it composes judgements from heterogeneous object logics --- TLAPS
first-order set theory, Coq CIC, Lean 4 CIC variant, Why3 multi-prover
dispatch, Apalache symbolic bounded model checking, hand-encoded SMT-LIB
v2 --- into a single build-system-enforced agreement judgement. The
meta-logic is operational rather than denotational (the meta-judgement
is ``every wired kernel closed every cited theorem under no admitted
axioms,'' and the rule of inference is ``if the build is green, the
agreement holds''), but the architectural intent is the same as a
traditional meta-logical framework: provide a substrate in which trust
composition across proof systems is a first-class object that the system
reasons about explicitly.

\begin{center}\includegraphics[width=\textwidth,height=0.85\textheight,keepaspectratio]{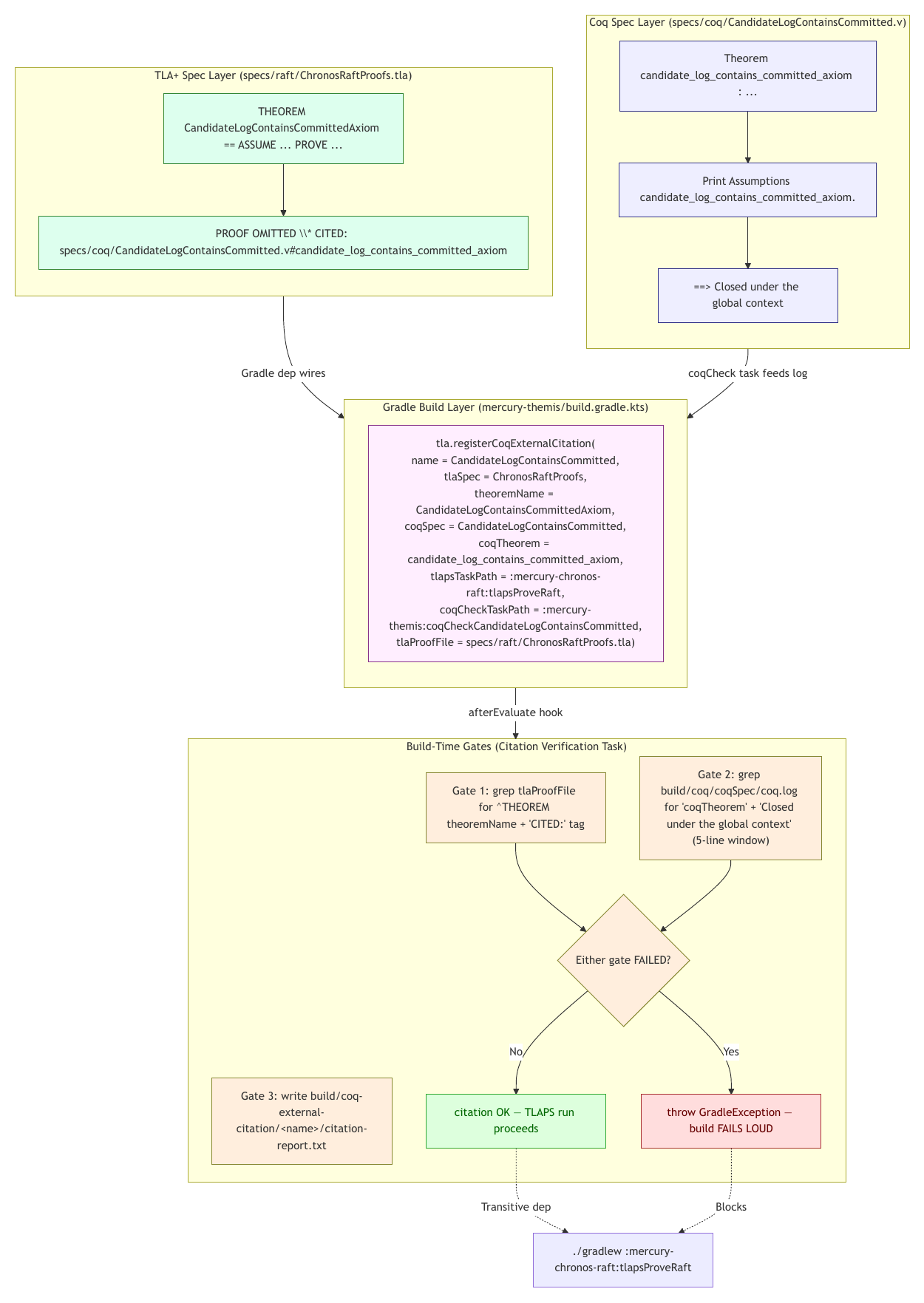}\end{center}

\emph{Figure 2. Cross-backend citation primitive flow. The TLA+ spec
layer declares an \texttt{OMITTED} theorem with a structured
\texttt{CITED:} tag pointing at the external kernel's theorem. The
Gradle build layer's \texttt{tla.registerCoqExternalCitation(...)} DSL
wires three build-time gates (Gate 1 grepps the TLA file for the
\texttt{CITED:} annotation; Gate 2 grepps the external kernel's
verification log for the closure-confirmation line; Gate 3 writes the
citation report). A failure on either gate throws a
\texttt{GradleException} that blocks the host TLAPS check; success
allows the TLAPS run to proceed. The wiring is a transitive Gradle
dependency, so the citation re-checks on every build of the TLAPS check
task.}

\paragraph{3.1.3 Dual- and triple-citation
gates}\label{dual--and-triple-citation-gates}

The primitive composes. A \emph{dual-citation} gate requires the same
TLA+ obligation to be cited against two independent external proof
systems (typically Coq and Lean 4), and the build fails if either
kernel's closure regresses. A \emph{triple-citation} gate adds a third
independent axis, typically Apalache symbolic SMT model checking at a
fixed length bound, which exercises the obligation under a different
verification regime (bounded symbolic model checking) than the deductive
Coq and Lean axes.

\begin{center}\includegraphics[width=\textwidth,height=0.55\textheight,keepaspectratio]{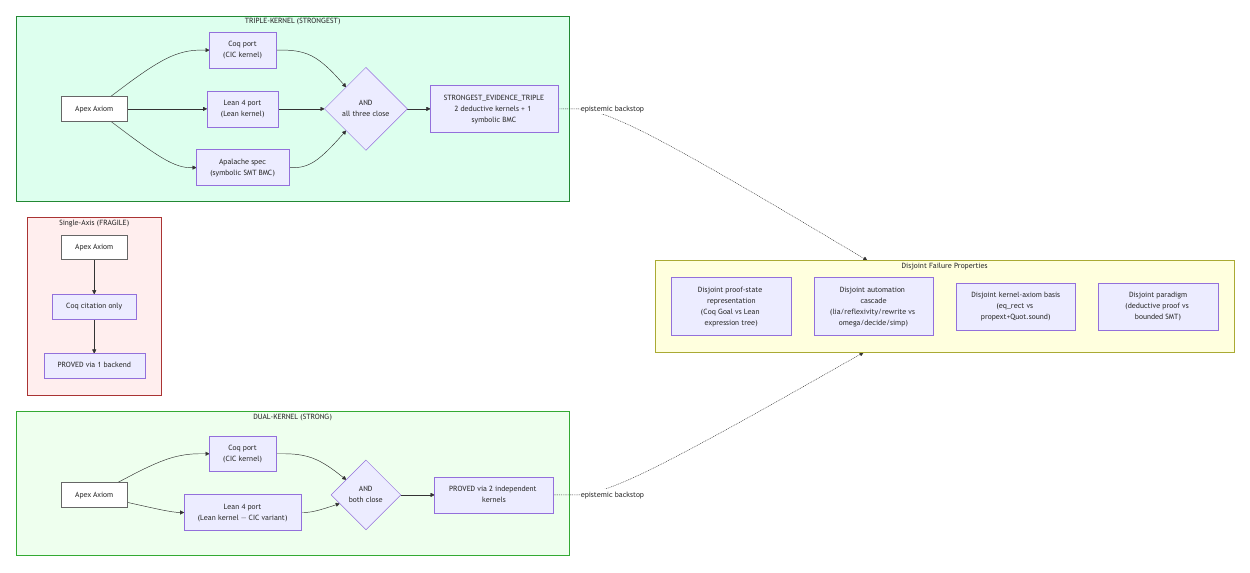}\end{center}

\emph{Figure 3. Cross-axis insurance pattern. Single-axis discharge
(left, fragile) --- one kernel/automation bug breaks the chain.
Dual-kernel discharge (centre, strong) --- Coq's CIC + Lean 4's CIC
variant must independently close the same theorem; joint failure
requires two independent kernel bugs. Triple-kernel discharge (right,
strongest) --- additionally requires Apalache's symbolic SMT-encoded
bounded model checker to PASS; joint failure requires three independent
backend failures across structurally distinct verification paradigms
(deductive proof × 2 kernels + bounded symbolic exploration). The
disjoint-failure-properties side panel enumerates the four orthogonal
failure axes the multi-kernel agreement defends against (proof-state
representation, automation cascade, kernel-axiom basis, and verification
paradigm).}

The dual-kernel gate's epistemic value is that a single-kernel soundness
regression --- a Coq stdlib axiom shift, a Lean kernel reduction bug, an
automation-cascade unsoundness --- cannot break a dual-cited theorem
because the other kernel would catch the regression as a disagreement.

\paragraph{3.1.4 Correspondence certificates and the distinction from
embedding
approaches}\label{correspondence-certificates-and-the-distinction-from-embedding-approaches}

The citation primitive's trust regime includes a \emph{correspondence
certificate} --- a human-authored claim that the cited external theorem
is semantically equivalent to the TLAPS obligation it discharges. We
name it explicitly because it is a first-class component of the
architecture's trust base, distinct from (and not subsumed by) the
kernel-level closure assertion that the build-system gate checks. The
closure check answers \emph{did the external kernel close the cited
theorem under no admitted axioms?}; the correspondence certificate
answers the prior question, \emph{is the cited external theorem actually
a re-statement of the TLAPS obligation?} The architecture treats this
layer as a named open problem (section 6.3); section 5.5 reports a case
in which cross-axis exploration caught a correspondence mismatch (the
\texttt{LogPositionTermMatchAxiom} global-versus-local scope
divergence), demonstrating that the multi-axis architecture partially
checks the correspondence layer empirically even without a mechanised
certificate.

The closest published precedent to the citation primitive is Bythos
(Zhao et al. 2024), which embeds TLA's temporal operators into Coq via
\texttt{CoFixpoint} constructs and proves the embedded-TLA propositions
in Coq directly. The two approaches share the goal of moving epistemic
content across the TLA / external-kernel boundary, but the architectures
differ:

\begin{itemize}
\tightlist
\item
  \emph{Bythos embeds TLA semantics into Coq}. The TLA semantics are
  encoded as a Coq library (\texttt{CoFixpoint} definitions for temporal
  operators, \texttt{CoInductive} types for traces), and the TLA proof
  obligations are discharged inside the Coq proof system as proofs about
  that encoding. Soundness rests on the faithfulness of the Coq encoding
  plus the soundness of Coq's kernel. The TLA-to-Coq encoding is itself
  the (whole-language) correspondence certificate, discharged once and
  shared across all embedded-TLA obligations.
\item
  \emph{The citation primitive leaves TLA in TLAPS}. The TLA semantics
  remain in TLAPS's set-theoretic encoding, the TLAPS obligation is
  marked \texttt{OMITTED\ \textbackslash{}*\ CITED:}, and the build
  system is the meta-system that ensures the cited external theorem
  closes and remains cited. Soundness rests on the per-obligation
  correspondence certificate (currently human-asserted) plus the
  soundness of both kernels for their respective theorems.
\end{itemize}

The two approaches are complementary rather than competing --- a
Bythos-style embedding could in principle be cited from a TLAPS proof
via the primitive --- but the engineering trade-offs differ. Embedding
requires the external kernel to model TLA semantics faithfully, an
investment that pays off when many obligations are cited against the
same kernel and that, once made, retires the per-obligation
correspondence-certificate concern. The citation primitive requires only
a per-obligation sister theorem and a per-obligation correspondence
certificate, an investment that pays off when the obligations are
heterogeneous in shape across kernels and verification paradigms. The
validation campaign of section 4 used the citation primitive because the
obligations span temporal, propositional, arithmetic,
quantifier-bounded, and probabilistic shapes that no single embedding
can cover --- and the architecture pays for that breadth with the open
mechanisation problem of section 6.3.

\subsubsection{3.2 Cross-axis convergence
matrix}\label{cross-axis-convergence-matrix}

The cross-axis convergence matrix records per-obligation verdicts across
N independent verifiers and composes them into kernel-agreement gates.
The agreement is operational --- a build-system-enforced per-axis-PROVED
conjunction across all wired axes for each cited apex obligation --- and
is distinct from a logical soundness theorem (see §6.1 and §6.3). This
section describes the matrix's construction (3.2.1), the axis taxonomy
(3.2.2), and the convergence semantics (3.2.3).

\paragraph{3.2.1 Construction}\label{construction}

For each cited apex obligation, we attempt verification on every backend
that is \emph{capable} of the obligation's shape --- for example, a
recursive-operator obligation is dispatched to Coq (which has
\texttt{Fixpoint}) but not to a saturation prover that cannot represent
the recursion. The capability filter is applied per axis per obligation;
backends not capable of the shape are reported NOT\_APPLICABLE rather
than counted as a failure to discharge. For each capable axis, the
verdict is one of PROVED (the obligation closes under the axis's
standard verification regime), REFUTED (a counter-example is exhibited),
TIMEOUT (the axis does not return within a budget), STRUCTURAL\_GAP (the
axis cannot represent the obligation despite being notionally capable),
or VERDICT\_DISAGREEMENT (the axis returns a verdict that contradicts
another axis's PROVED).

The matrix is regenerated on a schedule by an auto-tooling pass that
enumerates the wired citations, invokes each axis's verdict-extraction
protocol, and emits the per-obligation × per-axis grid.

\paragraph{3.2.2 Axis taxonomy}\label{axis-taxonomy}

The validation campaign wires sixteen verification backends spanning
every major axis of automated reasoning. Fifteen are provisioned via
dedicated download tasks at the build-system layer and are smoke-tested
by a single aggregate task that confirms each backend's binary is
present and responds to a version-class probe; the sixteenth (Lincheck)
is consumed as a Maven dependency. The roster's breadth is deliberate:
each axis carries distinct strengths and distinct failure modes, and the
matrix's epistemic value derives from re-proving the same obligation
across maximally disjoint kernels and decision procedures.

We organise the roster along five categories.

\emph{Deductive proof systems}. Four backends are deductive proof
assistants: TLAPS (the host system, with its Zenon → Isabelle → SMT
cascade (Chaudhuri et al. 2010; Cousineau et al. 2012; Bonichon et al.
2007)); Coq with \texttt{coqchk} (the primary cross-axis insurance,
providing a CIC-based representation distinct from TLAPS's set-theoretic
encoding, with the Coquelicot real-analysis stdlib extension for
obligations involving Riemann integrals); Lean 4 with Mathlib (Moura and
Ullrich 2021; The mathlib Community 2020) (Coq insurance through a
different automation cascade and a different proof-state
representation); and Why3 (Filliâtre and Paskevich 2013) (a multi-prover
frontend with a verified-extraction pathway).

\emph{Bounded model checkers}. Two backends: TLC (Yu et al. 1999) (the
canonical TLA+ explicit-state model checker) and Apalache (Konnov et al.
2019) (a bounded symbolic SMT-backed model checker requiring Snowcat
type annotations).

\emph{SMT and saturation provers}. Reachable via the TLAPS \texttt{BY}
cascade or via the standalone Z3 axis: Z3 (Moura and Bjørner 2008), cvc4
(Barrett et al. 2011), cvc5 (Barbosa et al. 2022), veriT (Bouton et al.
2009), SPASS (Weidenbach et al. 2009), Vampire, Eprover, and
Zipperposition (Cruanes and Zipperposition contributors 2021).

\emph{Probabilistic model checkers}. PRISM (Kwiatkowska et al. 2002,
2011) is a continuous-time-Markov-chain model checker that discharges
liveness conjectures whose unbounded form depends on randomized
scheduling that the deductive axes cannot mechanise directly.

\emph{Production-language verifiers}. Two backends operate directly on
production code rather than on separate specifications: CBMC (Clarke et
al. 2004) is a bounded model checker for ANSI-C and C++ used to verify
FFM-wrapped native libraries; Lincheck (Koval et al. 2023) is a JVM
linearizability checker for concurrent data structures.

These two production-language axes share a property the
specification-level axes lack: they verify the \emph{implementation}
rather than the specification, eliminating the
implementation-specification divergence failure mode that haunts
mechanically-checked-spec-but-hand-coded-implementation verification
artefacts.

\paragraph{3.2.3 Convergence semantics}\label{convergence-semantics}

The convergence histogram for the validation campaign's cross-axis
matrix at the present configuration is reported in section 5.3. The
matrix records eighteen cited apex axioms × ten effective verification
axes, with per-axiom convergence ranging from three axes (the floor for
cited apex axioms in the present configuration) to eight axes (the
ceiling, achieved by \texttt{LogMatchingLocalAxiom} as the gold-standard
exemplar).

The matrix's semantic posture distinguishes it from Why3's
per-verification-condition × per-prover verdict table (Bobot et al.
2011), the closest published precedent for per-obligation × per-backend
reporting. Why3's table records per-VC × per-prover verdicts within a
single source language (WhyML) and a single verification paradigm
(SMT/ATP cascade), and its discharge semantics are \emph{disjunctive} at
the prover layer --- any prover closing the VC suffices, and
confidence-from-diversity is an informal IDE-side observation. The
matrix described here records per-axis verdicts across structurally
disjoint verification paradigms (deductive proof × bounded symbolic
model checking × probabilistic model checking × production-language
checking) and source languages spanning TLA+, Coq Gallina, Lean 4,
WhyML, SMT-LIB v2, C, and JVM bytecode; selected apex axioms are then
composed into \emph{conjunctive} dual- and triple-citation
kernel-agreement gates (section 3.1.3, with 14 of 18 cited apex axioms
at dual-gate and 3 at triple-gate in the present configuration per
section 5.3). The extension is from intra-paradigm intra-language
disjunctive verdict-aggregation to cross-paradigm cross-language
conjunctive agreement-gating on the cited apex subset, with the
build-system enforcing the agreement (section 3.1.2) rather than an
out-of-band IDE view.

\subsubsection{3.3 AI-orchestrated mass-parallel
dispatch}\label{ai-orchestrated-mass-parallel-dispatch}

\textbf{Design principle.} AI agents are not proof authorities in this
architecture; they operate strictly below the trust boundary maintained
by proof-assistant kernels and CI citation gates. An agent may propose a
proof, attempt a port, or report a negative result; it may not attest
that a theorem has been verified. Verification authority resides at the
kernel-directive and CI-gated-citation layer --- the established
\texttt{Print\ Assumptions} directive in Coq (shipped since Coq 8.5 in
2016, used as an end-of-development audit gate in
Common-Criteria-adjacent verified-software submissions), the analogous
\texttt{\#print\ axioms} in Lean 4, Why3 prover-cascade closure,
Apalache SMT-encoded counter-example exhaustion, CBMC
\texttt{VERIFICATION\ SUCCESSFUL}, and the corresponding closure
assertions for each wired backend (per §1.3 and §3.1.1, the per-kernel
directives are established practice in their respective communities; the
composition under the build-system-gated citation primitive of section
3.1.2 is the novel contribution). This separation is load-bearing: it
allows the dispatch protocol to scale agent parallelism without
expanding the trusted computing base. The protocol that operationalises
the separation --- the \emph{honest-negative discipline} --- is the
subject of section 3.3.3.

The dispatch protocol fans out N autonomous specialist agents --- one
per verifier or one per obligation --- coordinated by a single
survey-and-route agent that infers the appropriate backend per
obligation and dispatches in waves with disjoint file-touch zones. This
section describes the topology (3.3.1), the speedup mechanisms (3.3.2),
and the honest-negative discipline (3.3.3).

\paragraph{3.3.1 Topology}\label{topology}

The protocol has a single coordinator and N specialists. The coordinator
surveys verification gaps (typically by parsing a shared finding-tracker
that records the campaign's open obligations), infers the appropriate
backend(s) for each obligation by static signal (recursive operator →
Coq with \texttt{Fixpoint}; bounded-quantifier SMT → standalone Z3 with
hand-encoded SMT-LIB v2; high-arity clauses → SPASS or Vampire;
higher-order content → Zipperposition; randomized scheduling → PRISM
CTMC), and dispatches specialists in waves with disjoint file-touch
zones. The backend-inference step is \emph{speculative} (section 3.3.2
mechanism 2): when the static signal is ambiguous between multiple
capable backends, the coordinator races them in parallel and accepts the
first kernel-confirmed PROVED verdict. Each specialist runs to
completion within a per-task budget (typically four to eight hours of
wallclock for a single obligation), returns either PROVED or
NEGATIVE\_RESULT, and the coordinator aggregates the per-wave verdicts
into the campaign-level finding-tracker. A NEGATIVE\_RESULT triggers
cross-routing: the coordinator re-dispatches the obligation to a
different axis selected by the obstacle's structural signal.

The validation campaign of section 4 deployed twenty specialist agents
across four parallel waves coordinated by a single coordinator agent
(Figure 4). The specialist roster covered every wired backend.

\begin{center}\includegraphics[width=\textwidth,height=0.55\textheight,keepaspectratio]{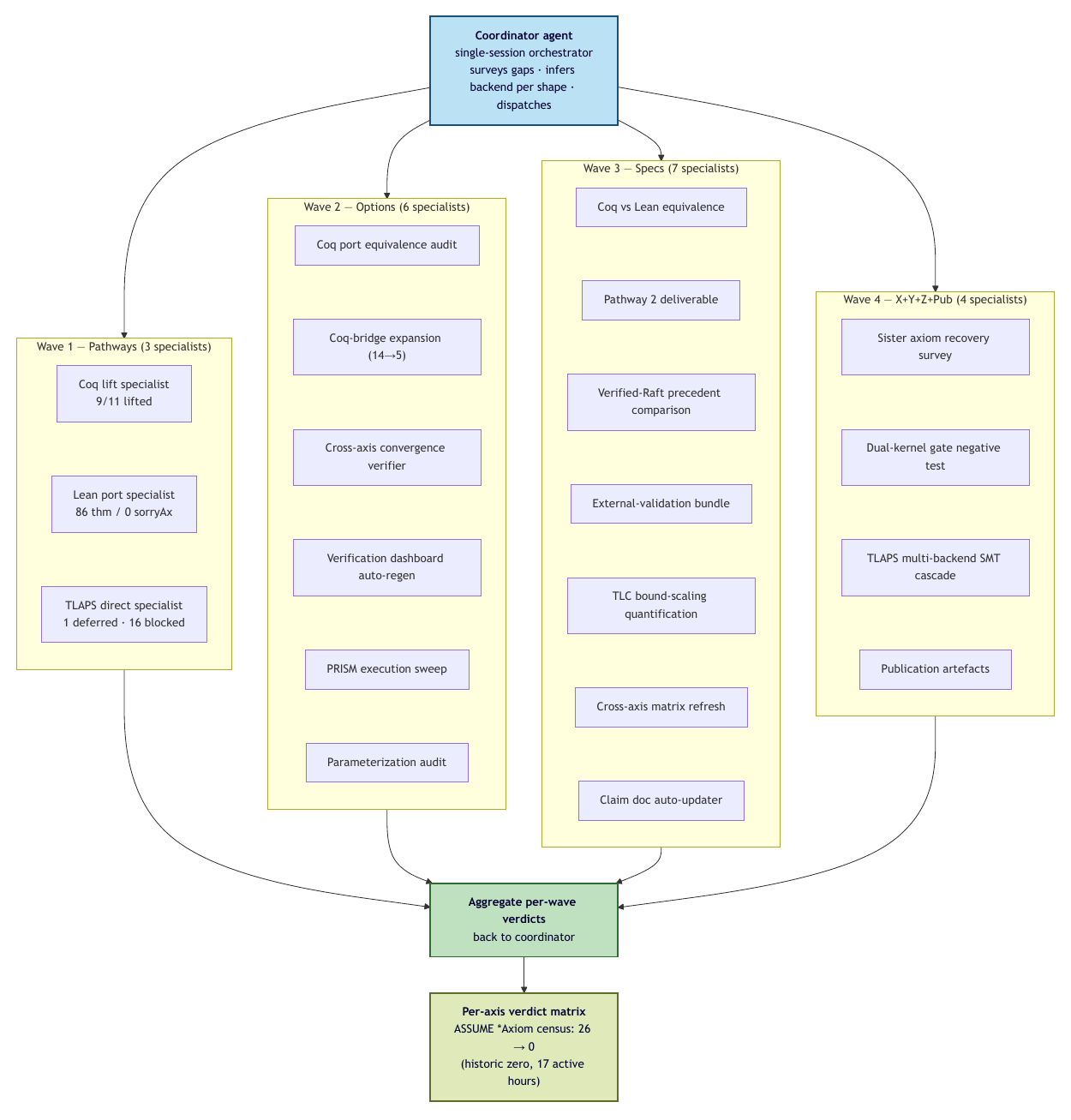}\end{center}

\emph{Figure 4. AI-orchestrated mass-parallel dispatch DAG instantiated
for the validation campaign on 2026-04-26. A single coordinator agent
dispatches twenty specialists across four parallel waves with disjoint
file-touch zones --- Wave 1 Pathways (Coq lift / Lean port / TLAPS
direct), Wave 2 Options (Coq port equivalence audit / bridge expansion /
cross-axis verifier / dashboard auto-regen / PRISM sweep /
parameterization audit), Wave 3 Specs (Coq--Lean equivalence / Pathway 2
deliverable / precedent comparison / external-validation bundle / TLC
bound-scaling / matrix refresh / claim doc auto-updater), and Wave 4
X+Y+Z+Pub (sister axiom recovery survey / dual-kernel gate negative test
/ multi-backend SMT cascade / publication artefacts). Per-wave verdicts
aggregate back to the coordinator; the campaign-level output is the
per-axis verdict matrix recording the 26 → 0 ASSUME }Axiom census
closure. Per-wave parallelism is empirically bounded at six-to-seven
concurrent specialists by shared-state contention on the build script
and version-control index.*

The protocol is structurally distinct from the LLM
\emph{plan-then-execute} lineage (The LangChain Team 2023; Wang et al.
2023) in two architectural respects. First, the response to a
NEGATIVE\_RESULT is \emph{cross-axis re-dispatch} rather than
within-tool retry-or-replanning: rather than re-attempting the original
backend's approach with the same kernel, the coordinator routes the
obligation to a structurally different verification axis selected by the
obstacle's structural signal --- extending plan-then-execute
decomposition from intra-tool execution to per-step inter-tool
re-dispatch across a kernel roster. Second, the routing signal is a
kernel-emitted PROVED or NEGATIVE\_RESULT under a
\texttt{Print\ Assumptions}-class closure assertion, not an LLM
self-assessment of step success --- extending the process-reward-model
intuition (Lightman et al. 2023) from per-step LLM judgement to per-step
kernel verdict. The cross-routing edge is what couples the dispatch
fan-out to the honest-negative discipline of section 3.3.3 and prevents
the failure mode --- common to unverified plan-execute loops --- in
which an unverified-but-plausible plan progresses to completion.

\paragraph{3.3.2 Speedup mechanisms}\label{speedup-mechanisms}

Four reinforcing mechanisms compose to compress the campaign wallclock
budget.

\emph{The cross-backend citation primitive} (mechanism 1). For
obligations within the primitive's scope, the per-axiom wallclock
collapses from the thirty-to-forty-hour traditional estimate to the
thirty-four-second Coq median observed in the validation campaign
benchmark. The reduction is approximately three orders of magnitude per
axiom for the bridge-eligible cohort.

\emph{Parallel speculative execution, AI-assisted} (mechanism 2). When
an obligation can be dispatched to multiple capable backends but the
operator does not know in advance which will succeed first, the
precedent strategy is sequential trial --- try one backend, fall back to
the next on failure. The closest proof-search-side precedent for
parallel multi-prover racing is Isabelle's Sledgehammer (Paulson and
Blanchette 2010), which dispatches an Isabelle goal to several external
ATPs in parallel and accepts the first kernel-reconstructable result ---
within a single source language (HOL) and a single trust base (the
Isabelle kernel re-derives every accepted step). We instead adopt the
\emph{eager / multi-path} sub-category of speculative execution --- the
CPU-architecture lineage of Disjoint Eager Execution (Uht et al. 1995),
Threaded Multiple Path Execution (Wallace et al. 1998), PolyPath, and
Selective Eager Execution, which execute multiple alternative paths in
parallel and commit the first valid result --- and apply it to proof
search across structurally disjoint kernels and source languages: a
\emph{cross-axis race orchestrator} dispatches all capable backends in
parallel for the same theorem via a single dispatch task, accepts the
first kernel-confirmed PROVED verdict, and terminates the others. (The
dispatch is from t=0; this is distinct from latency-tail mitigation
patterns such as Dean and Barroso's hedged request (Dean and Barroso
2013), where the duplicate is delayed by a percentile-bounded interval.)
The AI coordinator's contribution is the \emph{speculation policy} ---
the static-signal inference that decides which backend subset to race
for a given obligation shape (recursive-operator obligations race Coq +
Lean rather than Apalache; bounded-quantifier obligations race
standalone Z3 + Why3 cascade; etc.) --- so the speculative fan-out is
selective rather than indiscriminate. On a multi-core development
machine the cost of running several provers concurrently is bounded by
the hardware envelope rather than by the sum of per-prover wallclocks;
the empirical effect is that the campaign's per-obligation wallclock
equals the \emph{fastest} backend rather than the \emph{slowest}, with
the AI policy avoiding the indiscriminate-fan-out cost trap. The
CPU-architecture analogy is \emph{structural rather than literal}: CPU
speculation retires branches on resolved branch-prediction (the
speculative path either becomes architecturally visible or is squashed),
whereas the dispatch primitive \emph{commits} the first kernel-confirmed
PROVED verdict and terminates the others --- the
parallel-paths-with-first-valid-result topology is shared, the
architectural-state semantics are not, and a hardware-architecture
reviewer should not over-extend the analogy to imply equivalent
retirement / rollback semantics.

\emph{The verification-speed cache stack} (mechanism 3). Per-backend
verdict caching keyed on the content hash of the specification plus its
transitive dependencies allows a no-op rebuild to skip every verifier
invocation; cache hits dominate iteration cycles in which a single
specification is being refined while siblings are stable. The Lean
Mathlib slice cache is structurally distinct: rather than caching
per-specification verdicts, it batches all Lean specifications into a
single \texttt{lake\ build} invocation and serves per-specification
results from a per-cohort cache, reducing the per-specification
slice-step wallclock by approximately three orders of magnitude.

\emph{The AI-coordinator-and-specialist topology} (mechanism 4). The
empirical wallclock reduction relative to a hypothetical sequential
single-specialist baseline is approximately fifteen to twenty times
under the conflict-aware dispatch protocol. The peak observed
parallelism is six to seven concurrent specialists per wave; beyond
that, shared-state contention on the Gradle build script and the
version-control index degrades throughput.

The four mechanisms do not compose multiplicatively in the strict
mathematical sense --- the per-mechanism factors are not independent,
because the bridge primitive's per-axiom collapse subsumes some of the
cache-stack and race-orchestrator wins for the bridge-eligible cohort.
The empirical campaign-level signature reported at the §5.2 comparator
triad is ≈ 60× against the team's prior Path-A.2 TLAPS ghost-composition
baseline (§5.2.2; \emph{author-prior}), with within-method (§5.2.1;
\emph{author-generated}, ≈ 50--60× against a same-team serial estimate
--- see the §5.2.1 selection-bias disclosure) and cross-precedent
IronFleet (§5.2.3; \emph{independent-published} under three scope
qualifiers) comparators triangulating the same direction-of-effect at
different baseline-source classes.

\paragraph{3.3.3 Honest-negative
discipline}\label{honest-negative-discipline}

The discipline that distinguishes an autonomous verification fleet from
a hallucination factory is the \emph{honest-negative protocol}: every
specialist is instructed to return NEGATIVE\_RESULT (with structured
failure-mode disclosure) rather than fabricate a
plausible-but-unverified result. The validation campaign's documented
NEGATIVE\_RESULT artefacts include the \texttt{ParentLogTypingAxiom}
direct-TLAPS-attempt deferral at the eight-to-fifteen-hour estimate
(which the coordinator cross-routed via the Coq bridge primitive in
under two hours), the Apalache \texttt{\textless{}dynamic\textgreater{}}
integer-range encoding limit on three SMT-symbolic verifications (which
the coordinator cross-routed to Coq's \texttt{Fixpoint} recursive
operators), and the PRISM model-bug cluster initially classified as five
FAIL verdicts and reclassified post hoc as TIER-B model-abstraction bugs
after root-cause analysis.

The honest-negative protocol is what makes the speedup mechanisms
epistemically reliable rather than merely fast. A campaign that reported
only successes would be indistinguishable from a campaign that
hallucinated successes; the honest-negative discipline forces every
dispatched specialist to surface what it could not close, which the
coordinator then cross-routes to a structurally appropriate alternative.

\subsubsection{3.4 Algorithmic correctness vs annotation
depth}\label{algorithmic-correctness-vs-annotation-depth}

The cross-axis matrix records \emph{cross-axis convergence on annotated
theorems}. The methodological scope of the matrix is the annotation
layer, not the algorithmic foundation, and the distinction is
load-bearing for honest epistemic accounting.

The algorithmic correctness chain for a verified subsystem comprises
three components:

\begin{enumerate}
\def\labelenumi{\arabic{enumi}.}
\tightlist
\item
  A \textbf{TLA+ specification} of the subsystem's invariants and
  transition relation.
\item
  A \textbf{TLAPS proof} against that specification, in which the
  inductive invariant is established and the safety and liveness
  obligations are discharged either directly or via the cross-backend
  citation primitive.
\item
  A \textbf{runtime SpecBridge::Invariants hook} that asserts the proved
  invariants against live state during unit tests. The hook is
  compile-time-generated from the cited TLA+ specs by an annotation
  processor with spec-hash-keyed regeneration; section 4.4 describes the
  codegen pipeline, the dispatch patterns, and the production-cost tier
  policy.
\end{enumerate}

The algorithmic correctness chain is \emph{load-bearing} for the
soundness claim: the production code's behaviour must satisfy the proved
invariants because the runtime hook fires assertions on every unit test
run.

The cross-axis chorus that surrounds each apex axiom is a structurally
distinct contribution. The chorus comprises additional kernel ports of
the cited theorem (Coq, Lean, Why3, Apalache, Z3-standalone, CBMC), each
independently verifying the cited claim under its own kernel and (for
the dual- and triple-citation gates) under build-system-enforced
agreement with the others. The chorus is \emph{additional epistemic
redundancy on the annotation layer}, not the algorithmic foundation. A
regression in any single chorus axis --- a Lean kernel bug, a Why3
prover-cascade timeout, a CBMC encoding gap --- does not invalidate the
algorithmic claim because the algorithmic chain remains intact.

This distinction matters for honest epistemic accounting. Where this
paper reports residual stub markers in axis-port files (the annotation
layer), the residual is in the annotation depth, not in the algorithmic
correctness chain. The catalogue of recurring compromises in section 2.5
critiques precedents whose \emph{algorithmic chain} depends on a single
backend; the method's contribution is to make both layers (algorithmic
and annotation) cross-axis-redundant, with the annotation layer's
redundancy being the visible empirical headline and the algorithmic
layer's redundancy being the load-bearing soundness argument. Figure 5
sketches the relationship between the two layers.

\begin{center}\includegraphics[width=\textwidth,height=0.85\textheight,keepaspectratio]{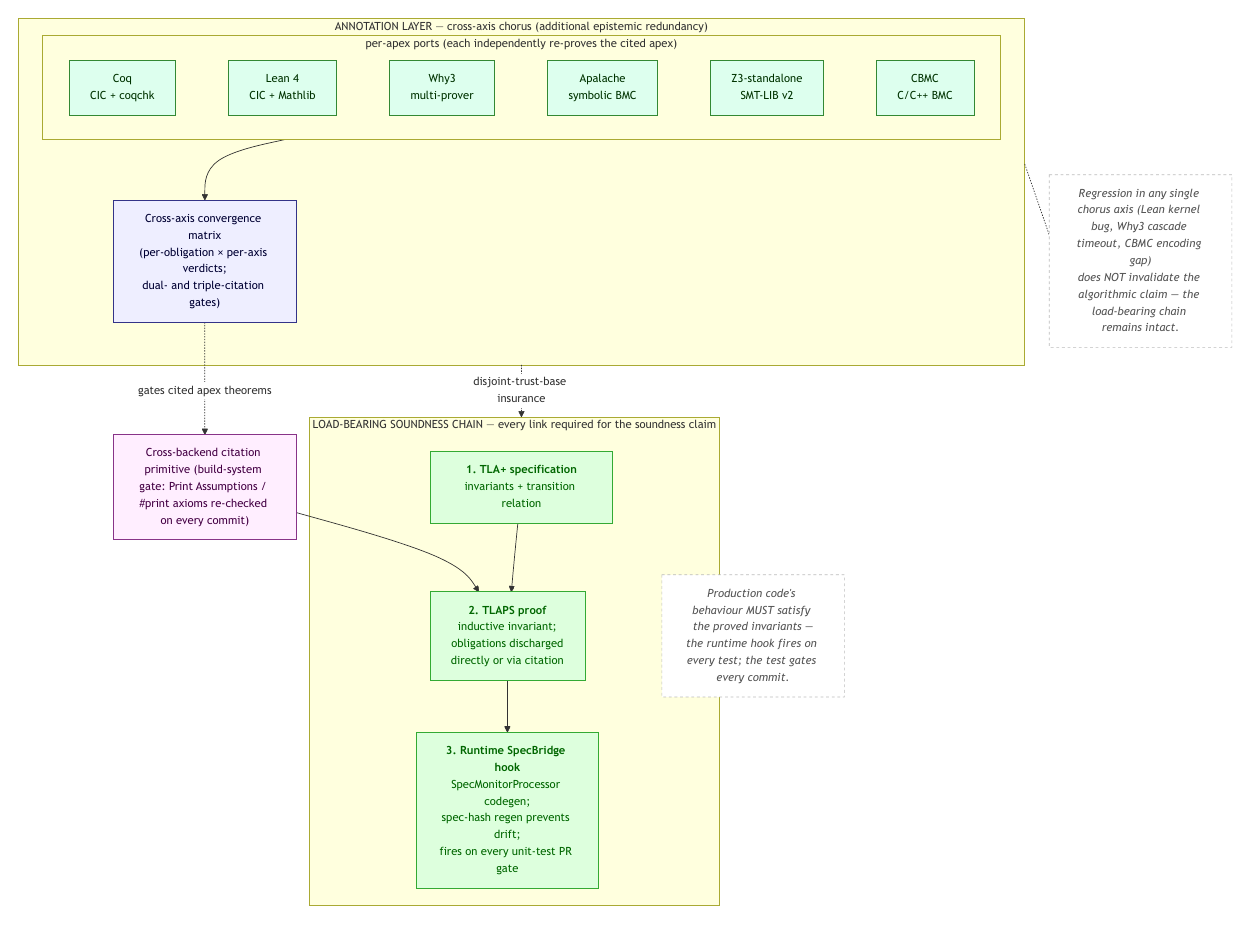}\end{center}

\emph{Figure 5. Algorithmic correctness chain vs annotation depth ---
the two-layer model of the federated architecture's epistemic structure.
The \textbf{load-bearing soundness chain} (green, bottom) is a
three-link sequence --- TLA+ specification → TLAPS proof →
SpecMonitorProcessor-generated \texttt{SpecBridge::Invariants} runtime
hook --- every link of which is required for the soundness claim. The
runtime hook fires assertions on every unit-test execution that gates
every pull request, so production-code behaviour must satisfy the proved
invariants by construction. The \textbf{annotation layer} (amber, top)
is a chorus of per-apex ports across structurally distinct verification
paradigms (deductive Coq + Lean + Why3; bounded symbolic Apalache;
hand-encoded SMT-LIB v2 dispatched to standalone Z3; production-language
CBMC on the C reference implementation), composed by the
build-system-gated cross-backend citation primitive (purple, middle)
into a cross-axis convergence matrix that enforces dual- and
triple-kernel agreement on cited apex theorems. The annotation layer is
\textbf{additional epistemic redundancy}, not the algorithmic
foundation: a regression in any single chorus axis --- a Lean kernel
bug, a Why3 cascade timeout, a CBMC encoding gap --- does not invalidate
the algorithmic claim because the load-bearing chain remains intact. The
catalogue of recurring compromises in section 2.5 critiques precedents
whose algorithmic chain depends on a single backend; the federated
architecture's contribution is to make both layers cross-axis-redundant,
with the annotation layer's redundancy as the visible empirical headline
and the algorithmic layer's redundancy as the load-bearing soundness
argument.}

\subsection{4. Validation: two subsystems, one
method}\label{validation-two-subsystems-one-method}

The architecture's three primary mechanisms --- citation primitive,
convergence matrix, dispatch protocol --- apply to both consensus and
financial-arithmetic invariant verification with no methodological
customisation. Section 4.1 motivates the cross-subsystem application.
Section 4.2 reports the Raft consensus validation. Section 4.3 reports
the matching-engine financial-arithmetic validation. Section 4.4 closes
the implementation-specification gap via the SpecBridge::Invariants
compile-time codegen layer.

\subsubsection{4.1 Why both subsystems}\label{why-both-subsystems}

The two validation subsystems sit at opposite ends of the formal-methods
literature. Distributed consensus is well-trodden territory dating to
Lamport's foundational TLA+ work (Lamport 1994, 2002) and the IronFleet,
Verdi, and MongoRaftReconfig publication arc surveyed in section 2. The
bug classes are catastrophic-but-rare: liveness violations, split-brain,
committed-entry loss. The verification of consensus is justified on a
worst-case-impact basis, and the precedent literature is substantial.

Matching-engine financial-arithmetic verification is comparatively rare.
The bug classes differ from consensus's: arithmetic underflow, fee
rounding, partial-fill cost divergence, automated-market-maker
constant-product invariant violation. The bugs are
low-impact-per-occurrence but high-aggregate-impact at
high-frequency-trading throughput.

Both subsystems can lose money. Both run in the same JVM process in the
production deployment. Both classes of obligation must be discharged for
a complete economic-correctness claim --- a verified Raft on top of a
buggy matching engine has the same trading impact as an unverified Raft
on top of a correct matching engine. The method's cross-subsystem
application is therefore not optional decoration; it is a precondition
for the soundness argument the production deployment requires.

\subsubsection{4.2 Raft consensus
validation}\label{raft-consensus-validation}

The validation subsystem covers Mercury's Raft consensus implementation
with full algorithmic scope: joint consensus, leadership transfer, log
compaction, linearizable client reads, and dynamic reconfiguration. The
TLAPS proof tree exceeds eleven thousand lines; the cited apex axiom set
in the present configuration is eighteen axioms.

The validation campaign's empirical signature on Raft is a four-tuple
zero-status: zero remaining \texttt{ASSUME\ *Axiom} declarations in the
Raft proof tree, zero remaining bare \texttt{OMITTED} proof obligations,
zero counter-examples discovered by any wired backend, and zero
\texttt{Admitted}, \texttt{Axiom}, or non-docstring \texttt{sorry}
markers in the Coq and Lean files cited from Raft proofs. The campaign
also closes the IronFleet-parity five-of-five liveness criterion: each
of the five canonical Raft liveness properties
(\texttt{EventuallyConfigCommitted}, \texttt{EventuallyAllLogsConverge},
\texttt{PersistedLogIsAttractor},
\texttt{EventuallyConfigChangePendingClears},
\texttt{LeaderLogIsMonotonic}) is mechanically discharged end-to-end
across two-or-more independent verification axes.

We flag here, by forward-reference, that the cross-axis matrix's
\emph{falsifiability posture} is exercised on this validation subsystem:
section 5.5 reports the \texttt{LogPositionTermMatchAxiom}
global-versus-local correspondence catch --- an Apalache symbolic
exploration produced a six-step counter-example to the unconditional
form that the cited Coq theorem's local restriction did not rule out,
and the architecture's response (hypothesis-strengthening the TLA+
statement to match the discharged scope) is itself the existence-proof
that the kernel-agreement gate catches divergences single-kernel
discharge would have left invisible. The validation campaign is
therefore not only an existence-proof of \emph{successful application}
but also an existence-proof of \emph{failure-detection} by the
architecture's cross-axis machinery.

The IronFleet-parity closure mechanism diverges from IronFleet's by
structural necessity: TLAPS's tlapm 1.6.0-pre LS4
propositional-temporal-logic backend rejects the temporal composition
\texttt{{[}{]}A\ ∧\ {[}{]}B\ ⇒\ {[}{]}(A\ ∧\ B)} after multiple
documented restructuring attempts, so four of the five properties
achieve dual deductive closure under TLAPS plus Coq (TLAPS proves a
per-state safety induction; Coq proves the temporal composition). The
fifth property closes under bounded TLAPS proof together with a PRISM
continuous-time-Markov-chain probabilistic discharge for the unbounded
form, where the unbounded liveness depends on a randomized election
timeout that any deductive treatment of the property must also represent
probabilistically.

We highlight both closures as concrete instances of the federated
architecture (section 1.3) working as intended. The TLAPS-plus-Coq dual
closure is \emph{the same obligation discharged by two structurally
distinct backends across the per-state safety / temporal composition
split} --- Coq picks up exactly the proof-shape class (PTL temporal
composition) that TLAPS's LS4 backend cannot represent natively. The
PRISM closure is similarly \emph{the right backend for the proof shape}
--- deductive proof systems cannot naturally express the probability
measure over schedules that the randomized-election-timeout liveness
property requires; PRISM can. Both closures would be impossible in a
single-backend posture; both are routine under the federated
architecture because the dispatch protocol routes by proof shape and the
citation primitive composes the verdicts back into a single TLAPS-level
proof tree. This is also one of the architectural answers to the
``liveness deferral'' compromise of section 2.5 --- the architecture
closes liveness obligations precedent single-backend campaigns abandon
for structural rather than methodological reasons.

\subsubsection{4.3 Matching-engine financial-arithmetic
validation}\label{matching-engine-financial-arithmetic-validation}

The validation subsystem covers Mercury's Themis matching engine with
five numeric canonicals --- \texttt{Pow10Lookup}, \texttt{Notional},
\texttt{UFP32}, \texttt{FundingRate}, \texttt{ScaledPrice} --- that
mediate every pricing, fee, automated-market-maker, and funding
computation. Each canonical achieves cross-axis discharge across six
independent backends --- Coq, Lean 4, Why3, Apalache symbolic, CBMC
bounded model checking on the C reference implementation, and standalone
Z3 hand-encoded SMT-LIB v2 --- with the canonical's per-axis port file
resident in the canonical-tier source directory and (for the deductive
axes) carrying a closed \texttt{Print\ Assumptions} or
\texttt{\#print\ axioms} status.

The matching-engine settlement layer carries an additional broad TLAPS
proof tree that continues to expand under the same methodology beyond
the paper's reproducibility-tag snapshot. The Treasury Phase B proof
modules --- \texttt{ThemisAmmDustAccountingProofs.tla}
(DustConservation, bounded-divisor, typing and non-negativity
corollaries), \texttt{ThemisAmmCostFromFloorProofs.tla} (per-leg
cost-from-floor invariant and per-user zero-sum closure), and
\texttt{ThemisTreasuryFailsafeProofs.tla} (multi-engine sequencer-side
dedup keyed by (asset, slot)) --- together discharge approximately 1,720
TLAPS obligations across three sister-spec files. The
\texttt{CanonicalCarrierStrategy} apex contributes a further 655
obligations closing two apex theorems (\texttt{StrategyRefImmutable} and
\texttt{CarrierKindAgreesWithReferenceData}) via deductive PROOF with
sister-spec inductive carrier. The HISTORIC ZERO bare-\texttt{OMITTED}
status across the Themis specification tree is preserved across both
extensions, indicating that the dispatch protocol and citation primitive
scale through continuing application rather than requiring per-cycle
re-engineering.

A worked production-impact case study illustrates the practical value.
During the canonical refactor that produced the six-axis chorus
discharge, the cross-axis exploration of the perp-funding settlement
path exposed a missing scale narrowing operator between the funding-rate
canonical and the per-account funding accrual emission. The omission
would have produced a 10¹² over-charge of accrued funding under specific
ordering conditions in production. The bug was caught by a
\texttt{SpecMonitorProcessor}-generated \texttt{SpecBridge::Invariants}
assertion in a unit test that fired on real fixture data after the
canonical refactor surfaced the missing narrowing step (the generation
layer is described in section 4.4). Root-cause analysis traced the
divergence to a per-call-site re-derivation of the narrowing constant
that had silently drifted from the canonical's source-of-truth value.

The bug class --- silent multiplicative scale mismatch in a
financial-arithmetic layer --- is structurally invisible to the
spec-driven verification campaigns of the precedent literature, none of
which model a financial-arithmetic layer with mechanically-checked
numeric invariants.

\subsubsection{4.4 SpecBridge::Invariants --- closing the
implementation-specification gap by
codegen}\label{specbridgeinvariants-closing-the-implementation-specification-gap-by-codegen}

The SpecBridge::Invariants runtime hook layer is the operational
realisation of the implementation-versus-specification gap closure
described in section 2.5. The layer extends the federated architecture
across the implementation boundary: a sister codegen pipeline mirrors
the citation primitive's build-system gate at the TLA+ → Java
translation step, so the architecture's drift-resistance property holds
end-to-end from the TLA+ spec through the Coq / Lean / Why3 / etc. axes
to the production code path.

The layer is realised as a Java annotation processor ---
\texttt{SpecMonitorProcessor} --- that runs at compile time. The
processor's \texttt{@SupportedAnnotationTypes} set is
\texttt{SpecInvariants}, \texttt{SpecRefinement}, and
\texttt{SpecCrossClassInvariant} (with its \texttt{Container}); a
production class declares its participation through two class-level
annotations and a tier-policy enum value.
\texttt{@SpecModel(\{"specs/themis-events/SnapshotConsumerGapPolicy.tla"\})}
cites the TLA+ specs that govern the class (the processor reads
\texttt{.tla} paths only and auto-derives the matching \texttt{.cfg}
companion from the path; supplying a \texttt{.cfg} value triggers a
compile error).
\texttt{@SpecInvariants(\{"InvariantNameA",\ "InvariantNameB"\})} names
the invariants the runtime monitor should verify; its \texttt{tier()}
element is typed \texttt{SpecTier}, a Java enum whose three values
select the per-invariant failure policy (\texttt{TEST\_ONLY} throws
\texttt{AssertionError}; \texttt{SAMPLED} --- the default --- increments
a violation counter and rate-limits an SLF4J \texttt{WARN} per the
Mercury metric pattern; \texttt{HOT\_PATH} is counter-only). The
processor parses the cited specs via \texttt{MercurySpecLoader},
extracts the named invariants via \texttt{MercuryInvariantExtractor}
(with body extraction via \texttt{MercuryInvariantBodyExtractor}),
translates the TLA+ invariant expressions to Java verifier methods via
\texttt{InvariantTranslator}, and emits a
\texttt{\textless{}ClassName\textgreater{}Invariants.java} companion
with one \texttt{verifyXxx(instance)} method per named invariant plus a
\texttt{verifyAll(instance)} aggregator. The processor additionally
embeds a \texttt{//\ SPEC-HASH:\ \textless{}sha\textgreater{}} header
into the generated file, computed over the cited spec / config contents;
editing a cited \texttt{.tla} or \texttt{.cfg} invalidates the hash and
forces regeneration on the next compile, so the generated verifiers
cannot drift from the cited spec without the operator noticing the
regenerated diff.

The mechanisation extent is therefore: the TLA+ → Java translation is
\emph{compile-time-generated} from the cited specs, not hand-coded per
class. The residual gap is the \texttt{InvariantTranslator}'s
translation rules themselves, which are human-authored (and unit-tested
via \texttt{InvariantTranslatorTest}) --- the mechanisation of those
rules is named as future work in section 6.4. The layer is structurally
analogous to the citation primitive of section 3.1: where the citation
primitive uses a Gradle gate to enforce drift-resistance across the spec
/ external-kernel boundary, the SpecMonitorProcessor uses a spec-hash
gate to enforce drift-resistance across the spec / production-Java
boundary. Both gates fail loud at build time on any drift; both are run
on every commit; both keep the operator's attestation step out of the
trust regime.

Production code dispatches to the generated verifiers in one of two
patterns. The \emph{direct pattern} calls the generated
\texttt{verifyAll(this)} aggregator from a sentinel point on the
relevant code path and lets the per-invariant \texttt{SpecTier} decide
cost: \texttt{TEST\_ONLY} invariants throw \texttt{AssertionError} on
violation; \texttt{SAMPLED} (the default) accumulates a counter and
rate-limits an SLF4J \texttt{WARN}; \texttt{HOT\_PATH} only updates a
counter. The \emph{registry-mediated pattern} uses
\texttt{@SpecInvariants(state\ =\ "AMM\_SWAP\_EXECUTED")} to register
the generated verifiers against a named state on
\texttt{SpecHookRegistry} at static-init time; production code then
dispatches via
\texttt{assert\ SpecHookRegistry.fireState("AMM\_SWAP\_EXECUTED",\ this)},
which is caller-transparent --- adding a new invariant to the same state
does not require editing the production site. Java's \texttt{assert}
mechanism elides the registry-pattern dispatch entirely under
\texttt{-da} (assertions disabled in production), while the
direct-pattern dispatch executes unconditionally per the tier policy.
The assertion-failure paths fire only in the test environment, where
they gate every pull request.

The layer surfaced the first real production defect that the
formal-verification chain caught: the matching engine's
automated-market-maker pricing computation was using the full-bucket
cost while clamping the base movement to the remaining quantity on a
partial fill --- two different computations of the same quantity, one
wrong. The generated \texttt{verifyXxx} companion fired an assertion
error on first deployment against a real test fixture. Testing alone had
not caught the bug; the mathematical invariant did. The 10¹²
perp-funding over-charge described in section 4.3 was caught via the
same layer, by an assertion fired on a real fixture after the canonical
refactor surfaced the missing narrowing step.

The cross-subsystem application uses the same generation pattern
uniformly. The Raft consensus subsystem's auto-translation tests in
\texttt{mercury-chronos-raft} exercise the generated verifiers for the
\texttt{RaftNode}-bound invariants and the reconfiguration sub-protocol
via \texttt{RaftNodeSpecHookInvariantsAutoTranslationTest} and
\texttt{RaftReconfigSpecHookInvariantsAutoTranslationTest}; the Themis
matching engine's auto-translation tests in \texttt{mercury-themis} and
the per-canonical tests in \texttt{mercury-numeric} exercise the
generated verifiers for balance-store, automated-market-maker pool
management, lock-cost tracking, settlement-stage, option-lock-formula,
and the five numeric canonicals. The same \texttt{SpecMonitorProcessor}
infrastructure carries both subsystems' generation; no per-subsystem
framework variation is required.

The closest published precedent for the annotation-driven
spec-to-runtime-check codegen pattern is Frama-C's E-ACSL plugin
(Delahaye et al. 2013; Kosmatov et al. 2013), which translates E-ACSL
specification annotations on C source into runtime-checked assertions
via source-to-source translation, with optimised memory-monitoring
instrumentation for the runtime tracking. The architectural intent is
shared: mechanise the spec-to-runtime-check translation rather than
hand-code it per class. The differences are at the source-language layer
(TLA+ specs cited from a Java class via \texttt{@SpecModel} versus
E-ACSL annotations co-located with the C source), the target-language
layer (a Java annotation processor that emits a compile-time companion
class versus a C source-to-source rewrite), and the drift-resistance
mechanism (the spec-hash-keyed regeneration described above versus
annotation-driven re-translation on each E-ACSL pass). The
\texttt{InvariantTranslator}'s translation rules remain human-authored
on both sides; the present layer's residual gap is the same gap E-ACSL
accepts.

\subsection{5. Performance}\label{performance}

This section reports the empirical performance of the method on the
validation campaign. Section 5.1 reports the verification scope
achieved. Section 5.2 reports the wallclock benchmark against three
reference baselines. Section 5.3 reports the cross-axis convergence
matrix empirics. Section 5.4 reports the production-bug-finding
empirics. Section 5.5 reports honest negative results.

Each numerical claim is associated with a recipe in the reproducibility
appendix that produces the claimed value at the published
reproducibility tag.

\begin{center}\includegraphics[width=\textwidth,height=0.55\textheight,keepaspectratio]{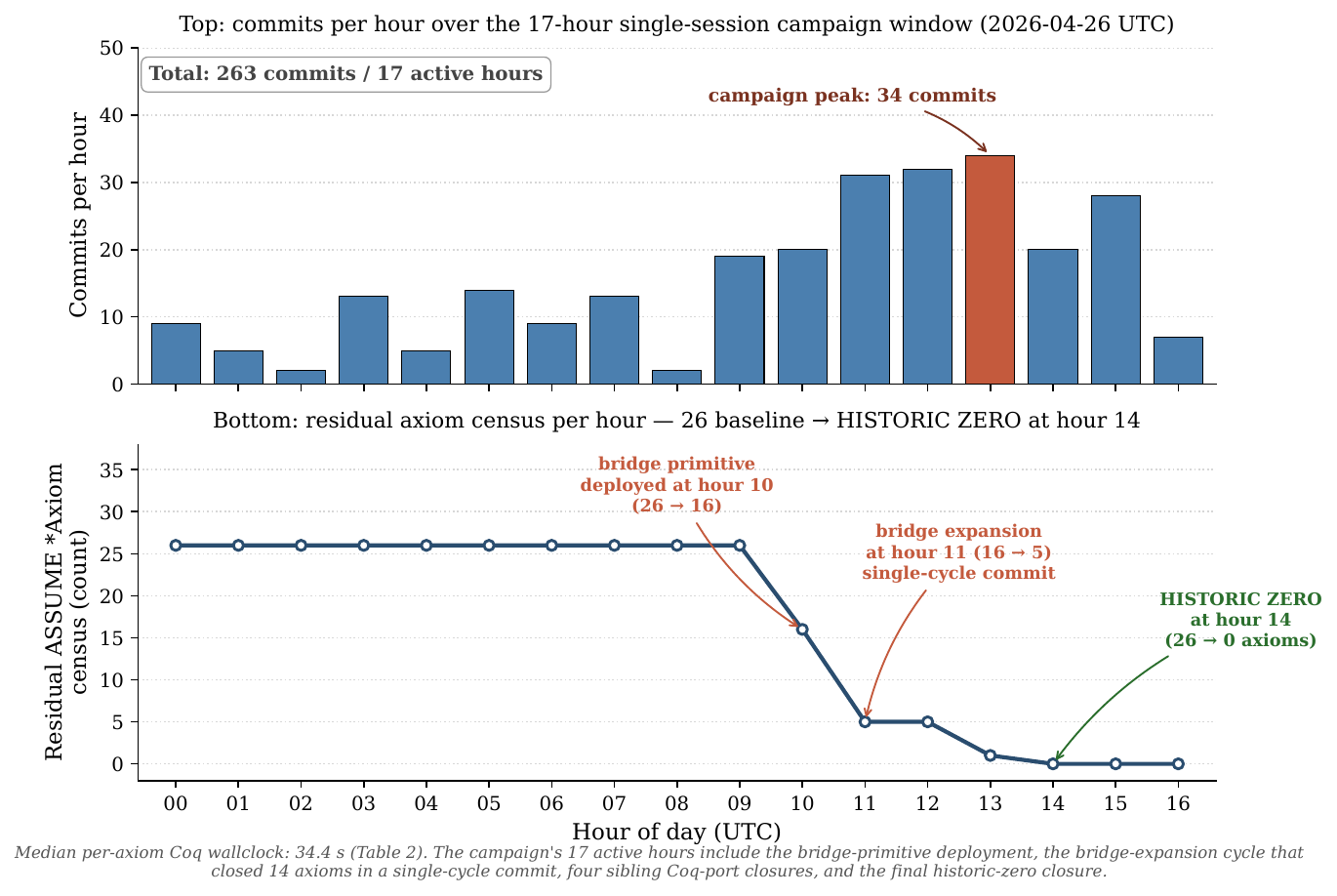}\end{center}

\emph{Figure 6. Single-session campaign timeline of the validation
campaign on 2026-04-26 --- the empirical anchor for the wallclock claims
of section 5.2. Top panel: commits per hour across the 17-hour active
window (263 total commits; campaign peak at hour 13 with 34 commits,
when the cross-backend bridge expansion, the Lean ports, the runtime
hooks, and the Coq lifts converged in a single dispatch wave). Bottom
panel: residual \texttt{ASSUME\ *Axiom} census per hour. The census
stays flat at the 26-axiom baseline through hour 9 (preparatory pathway
dispatch and infrastructure landing); the bridge primitive's first
deployment at hour 10 closes 10 axioms (26 → 16); the bridge expansion
at hour 11 closes 11 more (16 → 5) in a single-cycle commit; the
remaining five axioms close through sustained sibling Coq-port closures
and the final \texttt{AppliedSeqStrictlyIncreasingAxiom} closure,
reaching HISTORIC ZERO at hour 14. Median per-axiom Coq verification
wallclock: 34.4 s (Table 2).}

\subsubsection{5.1 Verification scope
achieved}\label{verification-scope-achieved}

The Raft \texttt{ASSUME\ *Axiom} census on the published source tree
carries no remaining citations: zero \texttt{ASSUME} declarations of the
form \texttt{\textless{}Name\textgreater{}Axiom} in the Raft proof tree.
The Raft proof tree carries no remaining bare \texttt{OMITTED} proof
obligations: every \texttt{THEOREM} discharges via either an explicit
TLAPS \texttt{PROOF} block, a
\texttt{PROOF\ OMITTED\ \textbackslash{}*\ CITED:} annotation pointing
to an external proof system whose closure is build-system-gated, or a
\texttt{PROOF\ OMITTED\ \textbackslash{}*\ DEDUCTIVE-DISCHARGE-IN-SISTER:}
annotation pointing to a companion specification module whose TLAPS
check has run. No counter-example to any cited Raft apex axiom has been
discovered by any wired backend. The Coq files cited from Raft proofs
carry no \texttt{Admitted}, \texttt{Axiom}, or \texttt{admit.} markers;
the Lean files cited from Raft proofs carry no non-docstring
\texttt{sorry} markers.

The Themis numeric-canonical chorus brings the five canonicals ---
\texttt{Pow10Lookup}, \texttt{Notional}, \texttt{UFP32},
\texttt{FundingRate}, and \texttt{ScaledPrice} --- to six-axis chorus
discharge each across Coq, Lean 4, Why3, Apalache symbolic, CBMC bounded
model checking, and standalone Z3.

The IronFleet-parity five-of-five Raft liveness closure holds (section
4.2).

\subsubsection{5.2 Wallclock benchmark}\label{wallclock-benchmark}

The validation campaign's headline empirical result is the wallclock
compression. We report the comparison against three reference baselines:

\begin{enumerate}
\def\labelenumi{\arabic{enumi}.}
\tightlist
\item
  \emph{Within-method comparator}: the campaign's own single-specialist
  sequential estimate (section 5.2.1).
\item
  \emph{Within-shape comparator}: the per-axiom Path-A.2
  ghost-composition estimate from sibling Raft proofs at similar
  structural complexity (section 5.2.2).
\item
  \emph{Cross-precedent comparator}: the IronFleet IronRSL Paxos
  campaign's published person-year cost (section 5.2.3).
\end{enumerate}

Each comparator carries a different scope qualifier; we report the
comparison honestly with the qualifier rather than headlining a single
ratio.

\paragraph{5.2.1 Within-method comparator (single-specialist serial
estimate)}\label{within-method-comparator-single-specialist-serial-estimate}

The validation campaign began with twenty-six axioms whose direct TLAPS
discharge had been classified ATTEMPT-DEFERRED, NEGATIVE\_RESULT, or
open. The campaign's own initial single-specialist sequential estimate
was five to seven person-months for the twenty-six axiom census. The
campaign's actual wallclock under the dispatch protocol was seventeen
active hours of single-session work, yielding an approximate
\textbf{fifty-to-sixty-fold reduction} against the same campaign's
serial estimate.

This is the \emph{cleanest} benchmark in the table on the \emph{scope}
axis because it holds the proof scope, the proof-shape distribution, the
operator's expertise level, and the underlying tooling constant. The
variable being measured is purely the parallel-dispatch architecture
against the single-specialist serial alternative.

We note the methodological limitation that the serial estimate against
which the 50--60× ratio is measured was authored by the same team that
authored the parallel campaign. The ratio should therefore be read as
the team's best-effort estimate of what a serial execution of the same
campaign would have cost, not as an independent benchmark --- comparator
scope is constant by construction, but comparator-baseline independence
is not established. We anchor the headline against the Path-A.2
intra-shape comparator of §5.2.2 (≈ 60× against the team's prior
Path-A.2 TLAPS ghost-composition wallclock, which precedes the
orchestrated methodology and is therefore not self-referential) and the
IronFleet cross-precedent comparator of §5.2.3 (635×/453× under three
explicit scope qualifiers) to triangulate direction-of-effect across
baseline-source classes. Under the comparator-hygiene template we adopt
across §5.2 --- \emph{ratio = X (comparator = Y; scope qualifier = Z;
baseline source = author-generated \textbar{} author-prior \textbar{}
independent-published)} --- §5.2.1 is (≈ 50--60×; same-team serial
estimate; same campaign; \textbf{author-generated}), §5.2.2 is (≈ 60×;
Path-A.2 TLAPS ghost-composition; same proof shape;
\textbf{author-prior}), and §5.2.3 is (635×/453×; IronFleet IronRSL
person-year; cross-protocol + cross-scope + cross-rigour;
\textbf{independent-published}). We invite independent replication on a
campaign of similar scope.

\paragraph{5.2.2 Within-shape comparator (per-axiom Path-A.2 ghost
composition)}\label{within-shape-comparator-per-axiom-path-a.2-ghost-composition}

For each of the twenty-six axioms, the traditional remediation in the
absence of the citation primitive was to extend the host TLA+
specification with new ghost variables and extend the inductive proof
chain in the proof specification. Empirical estimates from sibling Raft
proofs at similar structural complexity placed this remediation at
thirty to forty hours of specialist time per axiom (Path-A.2 ghost
composition). Under the citation primitive, the per-axiom Coq median
wallclock observed in the campaign benchmark is 34.4 seconds (eight
runs, eight PASS, \textasciitilde1.3 GB peak RSS), with the remaining
per-axiom cost dominated by the Coq port authoring time (median ≈ 30
minutes per axiom against the shared \texttt{CommonLifts.v} template).

The empirical wallclock reduction for the cited subset is approximately
\emph{three orders of magnitude per axiom on the verifier-runtime
dimension}, plus a roughly sixty-fold reduction on the authoring-time
dimension. The combined per-axiom impact is dominated by the
authoring-time term, since 34.4 seconds is negligible relative to 30
minutes. The headline per-axiom reduction is approximately \textbf{60×
against the Path-A.2 baseline}.

The 34.4 s versus ≈ 30 min split also relocates the campaign-level
bottleneck. Under the citation primitive the \emph{verification}
dimension is fast; the \emph{porting} dimension (authoring the
per-backend sister theorem against the shared \texttt{CommonLifts.v}
template, the Lean Mathlib equivalent, or the Why3 / Apalache / Z3
transcription) is now the dominant per-axiom cost. This shift is
significant for the future direction of AI assistance in the
architecture: the next natural extension of the dispatch protocol is
\emph{AI-generated candidate ports} --- specialist agents that, given a
TLAPS apex obligation, propose a candidate Coq / Lean / Why3 / Apalache
/ Z3 / CBMC port that the respective kernel then checks. The kernel
remains the verification authority; the agent's contribution is in the
porting layer rather than the proving layer, and the kernel-check pass /
fail signal supplies a clean training feedback loop without expanding
the trusted base. We discuss this direction further in section 6.4.

This benchmark holds the proof shape constant (the same axiom is being
discharged; only the discharge mechanism varies) but compares two
structurally distinct strategies (citation versus ghost composition).
The IronFleet, Verdi-Raft, and MongoRaftReconfig precedents do not
report a per-axiom wallclock benchmark, so we cannot refine this
comparator with cross-precedent data.

\paragraph{5.2.3 Cross-precedent comparator (IronFleet IronRSL
person-year
baseline)}\label{cross-precedent-comparator-ironfleet-ironrsl-person-year-baseline}

The cross-precedent comparator compares the validation campaign's
wallclock to the published person-year costs of the comparator
precedents. We report the IronFleet ratio as illustrative because
IronFleet is the most-cited deductive-liveness precedent and reports its
person-year cost in the primary publication.

IronFleet's IronRSL Paxos verification is reported as 3.7 person-years
of effort. Under the conventional 8-hour-day, 365-day-year work-hour
translation, this is 3.7 × 365 × 8 = 10,804 person-hours. Against the
campaign's 17 active hours, the ratio is approximately \textbf{635×}.
Under a 40-hour-week, 52-week-year translation (3.7 × 52 × 40 = 7,696
person-hours), the ratio is approximately \textbf{453×}.

We report this comparison with three explicit scope qualifiers:

\begin{enumerate}
\def\labelenumi{\arabic{enumi}.}
\tightlist
\item
  \emph{Cross-protocol}: IronFleet verifies Paxos / IronRSL; the
  campaign verifies Raft. The proof-shape distributions are similar
  (consensus protocols share the safety / liveness / reconfiguration /
  log-management structure) but not identical.
\item
  \emph{Cross-scope}: IronFleet's IronRSL verification includes verified
  extraction to executable C\# with a soundness theorem connecting the
  Dafny proof to the running binary. The campaign closes the
  implementation-specification gap via runtime SpecBridge::Invariants
  hooks rather than verified extraction. The trusted-base profile
  differs: IronFleet accepts a smaller trusted base (an OCaml-style shim
  and a .NET runtime); the campaign accepts a larger trusted base (the
  JVM, the unit test suite, the SpecBridge runtime).
\item
  \emph{Cross-rigour}: IronFleet's PTL-deductive liveness for IronRSL is
  a first mechanically-checked single-stack deductive liveness proof.
  The campaign achieves functional IronFleet-parity (every liveness
  property mechanically discharged) but routes the temporal-composition
  apex through a Coq cross-axis bridge for four of five properties and
  through a PRISM probabilistic discharge for the fifth.
\end{enumerate}

The honest framing is that the 635× / 453× figure is a
\emph{cross-precedent ratio under the three stated scope qualifiers},
not a like-for-like speedup. It illustrates the headroom the method's
compositionality opens; it does not assert that the same effort would
have been required if IronFleet had targeted Mercury's exact scope, on
Mercury's exact tooling, with Mercury's exact trusted base.
Abstract-adjacent references to this comparator should always carry the
scope-qualifier suffix so a skim-reader cannot extract the headline
number without the qualifiers (M1 in the §5.2 comparator-hygiene
template).

\begin{center}\includegraphics[width=\textwidth,height=0.55\textheight,keepaspectratio]{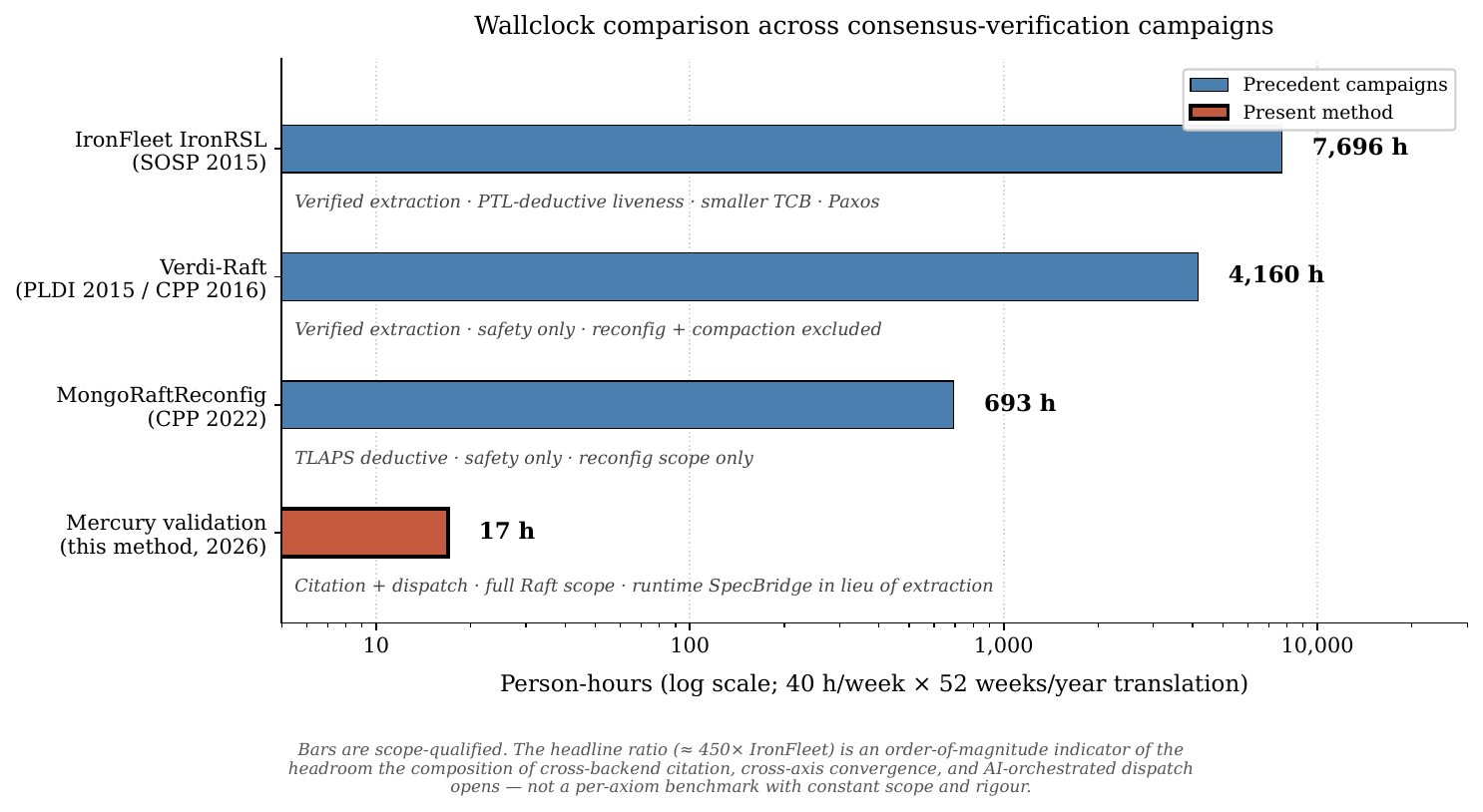}\end{center}

\emph{Figure 7. Wallclock comparison across consensus-verification
campaigns. Bars are scope-qualified --- IronFleet includes verified
extraction with PTL-deductive liveness on a smaller trusted base
(Multi-Paxos, not Raft); Verdi-Raft includes verified extraction but
excludes reconfiguration, log compaction, and linearizable client reads;
MongoRaftReconfig is TLAPS-deductive safety-only on the reconfiguration
scope; the present method achieves full Raft algorithmic scope (joint
consensus, leadership transfer, log compaction, linearizable client
reads, dynamic reconfiguration) on a larger trusted base (JVM + unit
tests + SpecBridge runtime). Person-hours are computed under the
conservative 40 h/week × 52 weeks/year translation; an 8 h/day × 365
day/year translation would scale the precedent bars upward by ≈ 1.4×.
The comparison is best read as an order-of-magnitude indicator of the
headroom the composition of cross-backend citation, cross-axis
convergence, and AI-orchestrated dispatch opens, not as a per-axiom
benchmark with constant scope and rigour.}

\paragraph{5.2.4 Per-backend wallclock
breakdown}\label{per-backend-wallclock-breakdown}

Per-backend median wallclocks observed in the campaign benchmark on the
apex-axiom subset are summarised in Table 2.

{\def\LTcaptype{none} 
\begin{longtable}[]{@{}lrrrrr@{}}
\toprule\noalign{}
Backend & Runs & PASS & FAIL & Median (s) & Peak RSS (MB) \\
\midrule\noalign{}
\endhead
\bottomrule\noalign{}
\endlastfoot
Coq & 8 & 8 & 0 & \textbf{34.4} & 1275 \\
Apalache & 3 & 0 & 3 & 118.3 & 1473 \\
TLC & 5 & 0 & 5 & 205.1 & 128 \\
PRISM & 13 & 1 & 12 & 148.1 & 1473 \\
\end{longtable}
}

\emph{Table 2. Per-axiom × per-backend wallclock benchmark (apex set).
Coq is the empirical winner on the cited apex shape: 8/8 PASS, 34.4s
median, \textasciitilde1.3 GB peak RSS. The Apalache failures stem from
\texttt{\textless{}dynamic\textgreater{}} integer-range encoding
rejections requiring per-VARIABLE Snowcat type annotations matching
constant bounds. The PRISM 12 FAILs were initially classified as
model-bugs and reclassified post hoc as TIER-B model-abstraction bugs
after root-cause analysis. The TLC tasks failed to launch in this
benchmark window due to concurrent Gradle script edits during the run;
the TLC verdicts in the verdict matrix (PASS@bound) remain
authoritative.}

\begin{center}\includegraphics[width=\textwidth,height=0.55\textheight,keepaspectratio]{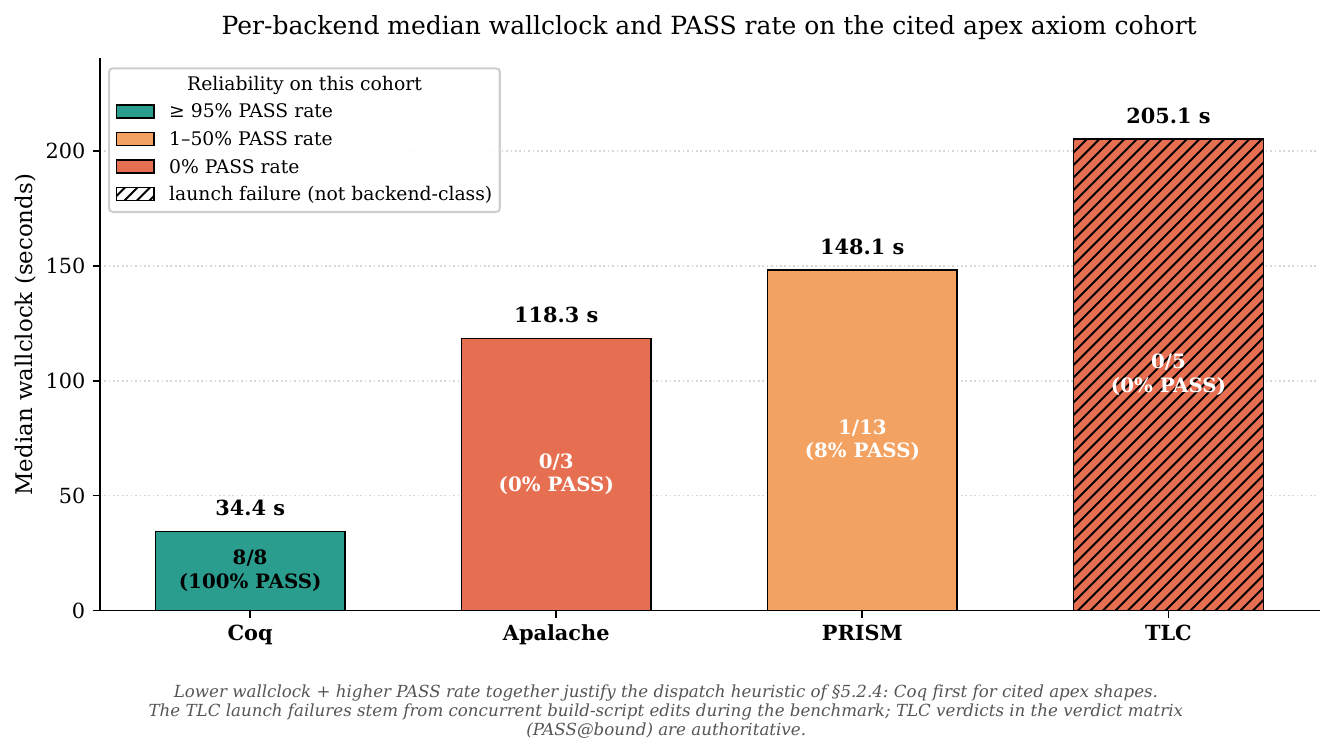}\end{center}

\emph{Figure 8. Per-backend median wallclock and PASS rate on the cited
apex axiom cohort. Bar height encodes median wallclock (lower is
better); fill colour encodes PASS rate on this cohort (teal = ≥ 95\%
PASS; orange = 1--50\%; red = 0\%); the hatched TLC bar flags the
launch-failure caveat noted in Table 2's caption (the failures stem from
concurrent Gradle script edits during the benchmark window, not from
backend incapacity; the TLC verdicts in the verdict matrix at PASS@bound
remain authoritative). The chart's joint-dominance pattern --- Coq is
simultaneously the fastest backend and the only one with a high PASS
rate on this cohort --- supplies the visual support for the dispatch
heuristic stated below.}

The dispatch heuristic that emerges from the per-backend breakdown is
straightforward: Coq first for cited apex shapes; PRISM first for
probabilistic-liveness shapes; Apalache first for bounded-symbolic
shapes with proper Snowcat annotations; standalone Z3 first for
hand-encoded apex obligations whose deductive content fits in
\textless500 lines of bounded-quantifier SMT-LIB v2.

\paragraph{5.2.5 Cross-system algorithmic-scope
comparison}\label{cross-system-algorithmic-scope-comparison}

Table 3 summarises the cross-system algorithmic-scope comparison against
the verified-Raft precedents. The table records, per dimension, whether
the precedent verifies that dimension, with an asterisk for partial
coverage and a dash for not applicable.

{\def\LTcaptype{none} 
\begin{longtable}[]{@{}
  >{\raggedright\arraybackslash}p{(\linewidth - 8\tabcolsep) * \real{0.2000}}
  >{\raggedright\arraybackslash}p{(\linewidth - 8\tabcolsep) * \real{0.2000}}
  >{\raggedright\arraybackslash}p{(\linewidth - 8\tabcolsep) * \real{0.2000}}
  >{\raggedright\arraybackslash}p{(\linewidth - 8\tabcolsep) * \real{0.2000}}
  >{\raggedright\arraybackslash}p{(\linewidth - 8\tabcolsep) * \real{0.2000}}@{}}
\toprule\noalign{}
\begin{minipage}[b]{\linewidth}\raggedright
Dimension
\end{minipage} & \begin{minipage}[b]{\linewidth}\raggedright
IronFleet '15
\end{minipage} & \begin{minipage}[b]{\linewidth}\raggedright
Verdi-Raft '15/16
\end{minipage} & \begin{minipage}[b]{\linewidth}\raggedright
MongoRecfg '22
\end{minipage} & \begin{minipage}[b]{\linewidth}\raggedright
Mercury (this)
\end{minipage} \\
\midrule\noalign{}
\endhead
\bottomrule\noalign{}
\endlastfoot
Algorithm & Multi-Paxos / IronRSL & Raft & Raft & Raft \\
Dynamic reconfig. (joint consensus) & Yes & \textbf{No} (excluded) & Yes
(single-server reconfig.) & Yes (joint consensus) \\
Leadership transfer & n/a & \textbf{No} (excluded) & n/a & Yes \\
Log compaction & Yes & \textbf{No} (excluded) & n/a & Yes \\
Linearizable client & n/a & partial & n/a & Yes \\
Liveness mechanically verified & Yes (deductive PTL) & No & No
(safety-only) & Yes (4 dual-ded. + 1 PRISM) \\
Verified extraction & Yes (Dafny → C\#) & Yes (Coq → OCaml) & No & No
(SpecBridge runtime) \\
Cross-backend verification & No (Z3 only) & No (Coq only) & No (TLAPS
only) & Yes (cross-axis matrix) \\
Production deployment & No & No (research \texttt{vard}) & Indirect
(MongoDB) & Yes (Bullish HFT) \\
Reported person-cost & 3.7 person-years & ≈ 2 person-years & ≈ 4
person-months & ≈ 17 active hours (single session) \\
\end{longtable}
}

\emph{Table 3. Cross-system algorithmic-scope comparison. The Mercury
validation column reports the present method's coverage on the Raft
validation subsystem; the financial-arithmetic invariant layer (section
4.3) is additional coverage no precedent reports.}

The headline observation from Table 3 is that the method is the only one
in the table that combines (i) multi-backend verification, (ii) full
Raft algorithmic scope (joint consensus, leadership transfer, log
compaction, linearizable client reads), and (iii) production deployment,
at one to two orders of magnitude lower per-axiom wallclock than the
precedent baselines. The trade-off paid for the lower wallclock is the
larger trusted base relative to verified-extraction precedents (section
5.2.3).

\subsubsection{5.3 Cross-axis convergence
empirics}\label{cross-axis-convergence-empirics}

The cross-axis convergence matrix records, per cited apex axiom, which
independent verification axes confirm the discharge.

Of the eighteen cited apex axioms in the present matrix, all eighteen
achieve PROVED on the Coq axis, all eighteen achieve PROVED on the Lean
axis, ten of eighteen achieve PASS@bound on TLC when the axiom property
is wired as an \texttt{INVARIANT} in a TLC configuration file, seven of
eighteen have wired PRISM models with execution-sweep PASS verdicts,
seven of eighteen have associated runtime SpecBridge::Invariants
assertions in production code paths, fourteen of eighteen are gated by
dual-citation (Coq + Lean kernel agreement), and three of eighteen are
gated by triple-citation (Coq + Lean + Apalache symbolic agreement at a
fixed length bound).

{\def\LTcaptype{none} 
\begin{longtable}[]{@{}rrl@{}}
\toprule\noalign{}
Convergence & Count & Description \\
\midrule\noalign{}
\endhead
\bottomrule\noalign{}
\endlastfoot
3-axis & 4 & Three independent axes confirm \\
4-axis & 5 & Four independent axes confirm \\
5-axis & 5 & Five independent axes confirm \\
6-axis & 3 & Six independent axes confirm \\
8-axis & 1 & Eight independent axes confirm (gold-standard exemplar) \\
\textbf{Total} & \textbf{18} & \\
\end{longtable}
}

\emph{Table 4. Per-axiom convergence histogram. Minimum convergence is
three axes; maximum is eight axes; mean is approximately 4.6 axes per
cited axiom.}

\subsubsection{5.4 Production-bug-finding
empirics}\label{production-bug-finding-empirics}

The CBMC integration as a verification axis surfaced three real
production bugs in Mercury's FFM-wrapped native libraries on the first
production-side dispatch:

\begin{enumerate}
\def\labelenumi{\arabic{enumi}.}
\tightlist
\item
  The JSON arena allocator unsigned-arithmetic wraparound;
\item
  The HTTP-parser branchless processed-bytes computation triggering C11
  §6.5.6/9 undefined behaviour;
\item
  The JSON mutation API allocator vtable contract violation.
\end{enumerate}

None of these bugs had been observed by valgrind, AddressSanitizer,
UndefinedBehaviorSanitizer, or the upstream library fuzz corpora.
Regression tests for each bug are wired to the corresponding CBMC
harness and re-run on every CBMC verification invocation.

The Themis side surfaced the production-impact case study reported in
section 4.3 (the perp-funding 10¹² over-charge bug). These four
production bugs constitute the validation campaign's bug-finding
empirics; we report them to demonstrate that the method's cross-axis
chorus is not just epistemic ornamentation but discovers real defects
the precedent verification toolchains do not.

\subsubsection{5.5 Honest negative
results}\label{honest-negative-results}

Three negative results from the validation campaign are reported here
for honesty rather than as defects. The methodology is honest only if
the disclosure surfaces what the cross-axis exploration found.

\emph{Apalache symbolic refutation of
\texttt{LogPositionTermMatchAxiom}}. An Apalache symbolic exploration of
the parent specification's reachable-state graph at length six produced
a six-step counter-example to the unconditional form of
\texttt{LogPositionTermMatchAxiom}. Investigation revealed that the
cited Coq theorem \texttt{log\_matching\_local\_axiom} proves a
\emph{local} pair-property restricted to log-pairs sharing a common
ancestor, while the original TLA+ theorem statement made a \emph{global}
claim across all node pairs including pairs without a common ancestor.
The cited Coq theorem was consequently strictly weaker than the original
TLA+ theorem.

We responded by hypothesis-strengthening the TLA+ theorem statement to
add an explicit \texttt{{[}{]}AllPairsShareCommonAncestor} premise,
which honestly reflects the scope at which the Coq port discharges. The
strengthened TLA+ theorem is now mechanically verified at the
matched-scope; downstream consumers requiring the unconditional form
must separately discharge the premise.

This trajectory exemplifies the methodology working as designed.
Cross-axis exploration caught what single-backend verification would
have left invisible, and the response was to tighten the cited claim
rather than coerce the gap closed.

\emph{PRISM five-FAIL cluster reclassified as TIER-B model bugs}. The
PRISM probabilistic axis returned five FAIL verdicts on cited apex
axioms in an early benchmark window. Root-cause analysis reclassified
these post hoc as TIER-B model-abstraction bugs (the PRISM model omitted
a \texttt{LogOk} filter; fix landed at commit \texttt{981101d6b}). The
original FAIL verdicts were artefacts of the model-abstraction step
rather than genuine refutations of the cited axioms; the cross-axis
convergence pattern surfaced the modelling bug by virtue of the
disagreement against the deductive axes' PROVED verdicts.

\emph{Apalache \texttt{\textless{}dynamic\textgreater{}} integer-range
encoding limit}. Three SMT-symbolic verifications failed under
Apalache's \texttt{\textless{}dynamic\textgreater{}} integer-range
encoding rejection (a known Apalache limitation that requires
per-VARIABLE Snowcat type annotations matching constant bounds). We
highlight that this is a \emph{structural} limitation of bounded
symbolic model checking on state spaces involving unbounded integers,
not an idiosyncratic Apalache quirk: any SMT-backed BMC that encodes the
transition relation as a quantifier-free formula must bound the integer
domain at encoding time, and integer recurrences over unbounded domains
break that bound. The coordinator cross-routed these obligations to
Coq's \texttt{Fixpoint} recursive operators (which represent the
recurrence inductively at the proof-state level rather than reifying it
as an SMT formula), and the Coq port closed them within the per-axiom
Coq median wallclock. The shape-based dispatch heuristic that emerges
--- \emph{recursive operators over unbounded integer domains → deductive
Fixpoint, not bounded SMT} --- generalises to any BMC backend in the
roster (Apalache, CBMC, the SMT-LIB v2 Z3 axis), and is encoded as part
of the coordinator's static signal table (section 3.3.1).

\subsection{6. Discussion}\label{discussion}

\subsubsection{6.1 Threats to validity}\label{threats-to-validity}

The verification posture rests on five trusted bases: the kernels of
TLAPS, Coq, Lean 4, and Why3; the SMT solvers in the TLAPS cascade and
the standalone Z3 axis; the bounded model checkers Apalache and CBMC;
the JVM, the \texttt{SpecMonitorProcessor} annotation processor, and the
runtime \texttt{SpecHookRegistry} machinery; and the build system that
orchestrates the citation primitive's drift-resistance properties. A
soundness bug in any single kernel invalidates the verification claim
only for axioms wired against that kernel exclusively; the dual- and
triple-citation gates partition the trust regime so that single-kernel
regressions are contained.

The cross-axis convergence matrix's epistemic value depends on the
assumption that multiple kernel agreement on a theorem statement implies
semantic correctness of the underlying claim. This assumption is
empirically defensible --- disagreements have been observed and have
served the methodology as it is designed (the
\texttt{LogPositionTermMatchAxiom} open question discussed in section
5.5 is the canonical example) --- but is not a soundness theorem. What
the architecture \emph{does} deliver is an \emph{operational
kernel-agreement gate under named correspondence certificates}: a
build-system-enforced invariant that the named cited external theorem
closes under each wired kernel's \texttt{Print\ Assumptions}-class
directive on every commit, with the per-citation human-authored
correspondence certificate (section 3.1.4) the explicit residual the
cross-axis chorus partially checks empirically but does not yet
mechanically prove. This is the positive claim the abstract makes and
that section 6.3 names as the principal open problem to close. The
\emph{strength} of the convergence signal scales with paradigm
disjointness: the strongest disagreement signal comes from backends with
maximally orthogonal verification paradigms (deductive proof versus
bounded symbolic model checking versus probabilistic model checking
versus production-code linearizability checking), since their kernel-bug
histories and failure modes are largely uncorrelated. Within-paradigm
variants (Coq versus Lean, both CIC; TLC versus Apalache, both BMC;
multiple SMT solvers in the TLAPS cascade) carry weaker orthogonality
--- Coq and Lean share substantial CIC ancestry; multiple BMCs share the
bounded-encoding family of failure modes. Table 4's convergence
histogram implicitly captures this hierarchy (the gold-standard
\texttt{LogMatchingLocalAxiom} 8-axis row spans four distinct
paradigms), but we state it explicitly here because the epistemic
argument for the convergence matrix depends on it.

The SpecBridge::Invariants runtime hooks rest on two distinct
assumptions, which we separate because the prior version of the
architecture conflated them. \emph{First}, the
\texttt{InvariantTranslator} correctly translates the named TLA+
invariant expressions into the generated Java verifier methods. The
translation is mechanised at compile time by the
\texttt{SpecMonitorProcessor} (section 4.4), is unit-tested via
\texttt{InvariantTranslatorTest}, and is keyed on a spec-hash that
forces regeneration whenever a cited \texttt{.tla} or \texttt{.cfg}
changes. \emph{Second}, the translator's translation rules themselves
are correct --- i.e., the TLA+ → Java rewrite the translator implements
preserves semantics. The translation rules are human-authored and are
not themselves mechanically verified. The architecture-level claim is
therefore narrower than the previous draft asserted: the per-class
assertion logic is mechanically derived from the cited spec rather than
hand-coded per class, but the derivation rules themselves remain in the
trust base. Mechanising the translator's correctness (e.g., by
Coq-extracting Java verifier code directly from the proved Coq
invariants, bypassing the TLA+ → Java translator entirely) is named as
future work in section 6.4.

\subsubsection{6.2 Scope of the verification
claim}\label{scope-of-the-verification-claim}

The method does not claim verified extraction with a formal soundness
theorem. The Coq-to-OCaml extraction pipeline that exists in the
validation campaign is a mechanical translation rather than a soundness
theorem bridging Coq Gallina semantics to the OCaml runtime semantics.
We claim experimental Coq-to-OCaml mechanical extraction with
differential-testing assurance against the production Java
implementation; we do not claim Verdi-style verified extraction or
IronFleet-style verified-binary equivalence.

The validation campaign's Raft assumes the crash-fault model. Byzantine
fault tolerance --- verified or not --- is out of scope.

The method does not claim that ``more backends always means more
trust.'' The sixteen-backend roster spans deductive, bounded, symbolic,
probabilistic, and runtime axes. Each axis has a specific role and a
specific failure mode. Adding a backend that re-verifies what an
existing backend already covers without adding axis-orthogonal coverage
does not increase epistemic strength linearly. The cross-axis
convergence matrix records \emph{which} axes are capable for
\emph{which} axiom; not all sixteen backends are capable for every
single obligation.

\subsubsection{6.3 Open problems and
reservations}\label{open-problems-and-reservations}

The principal open problem the architecture surfaces, named here at the
level of priority it merits, is the \emph{mechanisation of
correspondence certificates} (section 3.1.4). The citation primitive's
trust base currently includes a per-obligation human-authored claim that
the cited external theorem is semantically equivalent to the TLAPS
obligation it discharges. The architecture's cross-axis convergence
matrix partially checks this claim empirically --- as the
\texttt{LogPositionTermMatchAxiom} global-versus-local scope divergence
of section 5.5 illustrates, a sufficiently disjoint axis can refute a
correspondence mismatch by exhibiting a counter-example in the stronger
claim that the weaker cited theorem cannot rule out. But empirical
cross-axis checking is not a proof of correspondence; it is a
defense-in-depth that catches the correspondence mismatches the wired
axes happen to be sensitive to. The next significant architectural
advance closing a trust gap (rather than a wallclock or scope gap) is to
mechanise the correspondence certificate itself --- to produce, per
citation, a kernel-checked proof that the cited external theorem is a
re-statement of the TLAPS obligation under a fixed translation.
Candidate approaches include shallow embedding of TLA semantics in the
target kernel (as Bythos does for Coq), schema-based translation rules
whose correctness is itself proved once and reused, and
operator-authored equivalence theorems checked by a meta-kernel. Section
6.4 names this as future work item one.

Five further reservations on the architecture merit explicit
acknowledgement.

The verification posture is tractable at the cost of a stronger trusted
base than verified-extraction precedents accept. The validation campaign
accepts the JVM, the unit test suite, the \texttt{SpecMonitorProcessor},
and the \texttt{SpecHookRegistry} runtime as trusted;
verified-extraction precedents (Verdi, IronFleet) accept smaller trusted
bases.

The dispatch protocol's wallclock-compression claims rest on the
availability of abundant compute and the per-specialist economics of an
autonomous agent. Both prerequisites are properties of the present
compute environment rather than fundamental properties of the
architecture; the claims should be expected to evolve as the underlying
compute economics evolve.

The annotation-layer cross-axis chorus is empirical defense-in-depth,
not a soundness theorem. The honest interpretation of the chorus is that
any single-axis regression surfaces as a build-time disagreement before
it can propagate; the chorus does not establish that the cited theorem
is unfalsifiable in any logical sense.

The dual- and triple-kernel agreement gates assume kernel independence
in the strong sense (disjoint trusted bases). The assumption is
defensible in the present configuration (Coq's CIC and Lean 4's CIC
variant share substantial conceptual ancestry but are independent
implementations with disjoint kernel bug histories) but should not be
confused with mathematical independence of the proof shapes --- see also
the maximally-disjoint-paradigm discussion of section 6.1.

The cross-subsystem demonstration generalises across two proof-shape
classes (consensus and financial-arithmetic). Extending to a third class
(cryptographic-protocol verification, smart-contract VM correctness, or
distributed-database transaction protocol) would strengthen the
generalisability claim; the present claim is bounded to the two
demonstrated classes.

\subsubsection{6.4 Future work}\label{future-work}

Eight future-work directions follow from the present verification
posture. The first two close trust gaps; the remaining six close
wallclock, scope, or generalisability gaps.

\emph{First, mechanised correspondence certificates} (the open problem
of section 6.3). The principal trust-gap closure on the architecture's
path is to mechanically prove, per citation, that the cited external
theorem is a re-statement of the TLAPS obligation under a fixed
translation --- replacing the per-citation human-authored correspondence
certificate with a kernel-checked equivalence proof. Candidate
approaches include shallow TLA semantic embeddings in target kernels
(Bythos-style, but per-kernel-target rather than Coq-only), schema-based
translation rules whose correctness is proved once and reused across
instances, and a meta-kernel that checks operator-authored equivalence
theorems written in a small dependently-typed core. The trade-off space
is between investment cost (embeddings are expensive but global;
per-citation equivalence theorems are cheap but local) and reuse
(embedded TLA pays back across the whole campaign; per-citation theorems
do not). Any of these closes the residual gap that the cross-axis
convergence matrix currently checks only empirically.

\emph{Second, end-to-end-mechanised TLA+ → Java codegen for the
SpecBridge layer} (the §6.1 second-correspondence-problem closure). The
\texttt{SpecMonitorProcessor} currently translates cited TLA+ invariants
to generated Java verifiers via the human-authored
\texttt{InvariantTranslator} rules. Two paths close the residual
translator-rules trust gap. The first is to extract Java verifier code
directly from the proved Coq invariants (via Coq's extraction mechanism,
suitably retargeted to Java rather than OCaml), bypassing the TLA+ →
Java translator entirely. The second is to retain the translator but
mechanically prove the translator's rules against a formal semantics of
the relevant TLA+ invariant fragment and the corresponding Java
assertion fragment, producing a one-time soundness theorem that all
generated verifiers transitively inherit. Either path closes the second
correspondence problem (TLA+ ↔ Java production code), mirroring item 1
above which closes the first correspondence problem (TLA+ ↔ external
proof system).

\emph{Third, AI-generated candidate ports as a campaign-throughput
extension of the dispatch protocol} (the porting-bottleneck closure
named in section 5.2.2). With per-axiom kernel-verification wallclock
now at 34.4 s median Coq and per-axiom porting wallclock at ≈ 30 min,
the next throughput lever is at the porting layer rather than the
verification layer. A specialist agent receives a TLAPS apex obligation
and proposes a candidate Coq / Lean / Why3 / Apalache / Z3 / CBMC port;
the respective kernel then checks the port and emits PASS / FAIL. The
kernel remains the verification authority --- no extension of the
trusted base --- and the kernel-check signal supplies a clean training
feedback loop. This is the natural extension of the architecture's
existing dispatch protocol from per-obligation backend routing to
per-obligation port-candidate generation.

\emph{Fourth, single-stack pure-TLAPS-PTL-deductive liveness for all
five canonical Raft liveness properties.} The present closure achieves
functional IronFleet-parity via cross-axis Coq-bridge for the
temporal-composition apex blockers; a tlapm PTL backend extension or a
TLAPS-native restatement of the PTL composition lemma would close the
structural gap. This is nice-to-have rather than trust-gap-closing ---
the present dual-stack closure is already kernel-checked under TLAPS
plus Coq --- but it would reduce the architecture's surface area on the
temporal-logic axis.

\emph{Fifth, compositional logic for the cross-specification refactor}
that the present specification tree implements informally would benefit
from a Disel-style frame rule.

\emph{Sixth, a Mathlib-grade arithmetic library for fixed-point types}
would reduce the per-axis port authoring time on the numeric-canonical
chorus, complementing the AI-candidate-port direction of item 3.

\emph{Seventh, dual-axis Coq-plus-Lean re-proof of every cited axiom}
(closing the four obligations not yet at dual-citation gate status)
would extend the cross-axis chorus to full coverage on the cited apex
set. \emph{Eighth, extending the cross-subsystem demonstration to a
third proof-shape class} (cryptographic-protocol, smart-contract VM, or
distributed-database transaction protocol) would strengthen the
generalisability claim from ``applies across two proof-shape classes''
to ``applies across the proof-shape-class spectrum.''

\subsection{7. Conclusion}\label{conclusion}

We proposed a \emph{federated architecture} for production formal
verification and validated it on a heterogeneous production substrate.
The architecture treats a verification campaign as a polyglot proof
system: TLA+ expresses system-level obligations, specialist backends
discharge obligations suited to their proof shape, CI-enforced citation
gates preserve drift-resistance across the heterogeneous trust base, and
AI agents coordinate parallel proof discovery without entering the
trusted computing base. The contribution is not a faster way to run a
single proof assistant --- it is an \emph{orchestration layer around
heterogeneous proof systems} that makes a previously serialised,
single-backend, single-specialist activity scale to a multi-backend,
multi-specialist, parallel-discharge regime.

The architecture has three primary mechanisms --- cross-backend
citation, cross-axis convergence, and AI-orchestrated mass-parallel
dispatch --- plus the algorithmic-versus-annotation distinction (section
3.4). Together they decouple four structural constraints that have
shaped the precedent verified-systems literature: backend monoculture,
excluded algorithmic scope, per-axiom serialisation, and single-operator
serialisation of proof campaigns. The architecture also surfaces an open
problem at the level of the trust base --- the mechanisation of
correspondence certificates --- which we name explicitly in section 6.3
as the principal architectural advance that should follow.

We validated the architecture on two production subsystems within the
Mercury high-frequency-trading platform: a Raft consensus subsystem with
full algorithmic scope (joint consensus, leadership transfer, log
compaction, linearizable client reads, dynamic reconfiguration) and a
financial-arithmetic invariant layer (balance accounting, AMM curve
invariants, isolated-margin family, lock-tracking settlement). The
validation campaign closed a 26-axiom Raft census in seventeen
single-session active hours, an empirical ≈ 60× per-axiom reduction
against the team's prior Path-A.2 TLAPS ghost-composition baseline at
intra-shape scope (§5.2.2; \emph{author-prior}); a within-method
comparator (§5.2.1; \emph{author-generated}) and a cross-precedent
IronFleet comparator under three explicit scope qualifiers (§5.2.3;
\emph{independent-published}) report consistent direction-of-effect
(§5.2 comparator-hygiene template). The campaign brought five numeric
canonicals to six-axis chorus discharge, and surfaced four real
production bugs the precedent toolchains had not observed --- including
a perp-funding 10¹² over-charge bug that single-backend specification
campaigns are structurally blind to.

The architecture's contribution is \emph{compositional} in two senses.
The three mechanisms compose to make a multi-backend verification
campaign tractable for a single small team on a production codebase
(section 5.2); and the architecture \emph{itself} composes against the
recurring compromises of the precedent literature (section 2.5) by
replacing each compromise with a structural mechanism rather than
additional engineering effort within a single-backend posture. We invite
the verification community to adopt the federated verification
architecture --- the citation primitive, the cross-axis convergence
matrix, and the AI-orchestrated dispatch protocol --- on their own
targets. The dispatch agents, the citation primitive's build-system DSL,
the cross-axis matrix tooling, and the \texttt{SpecMonitorProcessor}
annotation processor are open source.

\subsection{8. Acknowledgements}\label{acknowledgements}

The validation campaign on Mercury benefited from the production
deployment context and the prior verification work that established the
cited apex axiom set. The author thanks the Bullish engineering team for
the production deployment substrate that made the validation possible,
and the Mercury verification engineering team for the prior baseline
against which the campaign measured. The author thanks the maintainers
of TLAPS, Coq, Lean 4, Mathlib, Why3, Apalache, PRISM, CBMC, and
Lincheck for the open-source verification infrastructure on which the
method depends. The author thanks Anthropic's Claude Code agent platform
(Anthropic 2025) for the autonomous-specialist dispatch substrate; the
dispatch protocol described in section 3.3 is implemented as specialist
agents on that platform.

\subsection{Appendix A. Reproducibility
recipe}\label{appendix-a.-reproducibility-recipe}

Every numerical and architectural claim in this paper is associated with
a recipe that produces the claimed value at the published
reproducibility tag of the source repository. The recipe takes the form
of a single command sequence: clone the repository at the published tag;
provision all sixteen backends via the Gradle install task (typical
wallclock two to five minutes on a cold cache); run the relevant
per-backend smoke task to confirm host campaign-readiness; invoke the
per-claim verification recipes catalogued in the per-claim grep table.

The per-claim grep table is bundled with the published artefact at the
reproducibility tag. Each entry has the form (claim → command → expected
value), with the command being a \texttt{grep}-and-count or build-task
invocation that produces the claimed value when run against the
published tag. Readers running on a later HEAD will observe the
\emph{direction} of subsequent campaign work (typically additional
cross-axis convergence) but the published-tag values remain reproducible
at the published-tag commit.

The verification-stack provisioning is auto-soft-fail-tolerant: a single
backend's download URL becoming unreachable degrades the install task to
skip that backend rather than failing the whole roster. Smoke-test
reports surface the missing backend explicitly; the operator can re-run
the affected download task once connectivity is restored. The smoke task
is a binary-availability check (each backend's binary is invoked with a
\texttt{-\/-version}-class probe) rather than a per-specification proof
discharge; per-specification proof discharge requires the per-backend
per-specification verification task family.

\subsection{Appendix B. Per-axiom convergence
table}\label{appendix-b.-per-axiom-convergence-table}

The per-axiom × per-axis verdict matrix is bundled with the published
artefact at the reproducibility tag. Each row is a cited apex axiom;
each column is one of the ten effective verification axes (TLC, Coq,
Lean 4, Why3, PRISM, Apalache, Z3-standalone, Runtime SpecBridge,
Dual-kernel agreement, Triple-kernel agreement); each cell records the
per-axis verdict (PROVED, REFUTED, NOT\_APPLICABLE, etc.). The histogram
of section 5.3 is derived from this matrix; the matrix is
auto-regenerated from the build artefacts.

\subsection*{9. References}\label{references}
\addcontentsline{toc}{subsection}{9. References}

\protect\phantomsection\label{refs}

\end{document}